\documentclass[12pt,preprint]{aastex}

\def\arcsecpoint{$''\!.$}
\def\deg{$^{\rm o}$}

\def\gtsim{\raisebox{-.5ex}{$\;\stackrel{>}{\sim}\;$}}

\shortauthors{Crenshaw et al.}
\shorttitle{Mass Outflow in NGC~5548}

\begin{document}

\title{Mass Outflow in the Seyfert 1 Galaxy NGC~5548\altaffilmark{1}}

\author{D.M. Crenshaw\altaffilmark{2},
S.B. Kraemer\altaffilmark{3},
H.R. Schmitt\altaffilmark{4},
J.S. Kaastra\altaffilmark{5,6},
N. Arav\altaffilmark{7},
J.R. Gabel\altaffilmark{8},
and K.T. Korista\altaffilmark{9}}

\altaffiltext{1}{Based on observations made with the NASA/ESA Hubble Space 
Telescope, obtained at the Space Telescope Science Institute, which is 
operated by the Association of Universities for Research in Astronomy, Inc., 
under NASA contract NAS 5-26555. These observations are associated with 
proposal 9511.}

\altaffiltext{2}{Department of Physics and Astronomy, Georgia State 
University, Astronomy Offices, One Park Place South SE, Suite 700,
Atlanta, GA 30303; crenshaw@chara.gsu.edu}

\altaffiltext{3}{Institute for Astrophysics and Computational Sciences,
Department of Physics, The Catholic University of America, Washington, DC
20064}

\altaffiltext{4}{Remote Sensing Division, Naval Research Laboratory,
Washington, DC 20375; and Interferometrics, Inc., Herndon, VA 20171}

\altaffiltext{5}{SRON Netherlands Institute for Space Research,
Sorbonnelaan 2, 3584 CA Utrecht, The Netherlands}

\altaffiltext{6}{Sterrenkundig Instituut, Universiteit Utrecht, 
P.O. Box 80000, 3508 TA Utrecht, The Netherlands}

\altaffiltext{7}{Physics Department, Virginia Polytechnic Institute \&
State University, Blacksburg, VA 24061}

\altaffiltext{8}{Physics Department, Creighton University, Omaha NE 68178}

\altaffiltext{9}{Department of Physics, Western Michigan University,
Kalamazoo, MI 49008}

\begin{abstract}
We present a study of the intrinsic UV absorption and emission lines in an
historically low-state spectrum of the Seyfert 1 galaxy NGC~5548, which we
obtained in 2004 February at high spatial and spectral resolution with the
Space Telescope Imaging Spectrograph (STIS) on the {\it Hubble Space
Telescope} ({\it HST}). We isolate a component of emission with a width of
680 km s$^{-1}$ (FWHM) that arises from an ``intermediate line region''
(ILR), similar to the one we discovered in NGC~4151, at a distance of
$\sim$1 pc from the central continuum source. From a detailed analysis of
the five intrinsic absorption components in NGC~5548 and their behavior
over a span of 8 years, we present evidence that most of the UV absorbers
only partially cover the ILR and do not cover an extended region of UV
continuum emission, most likely from hot stars in the circumnuclear
region. We also find that four of the UV absorbers are at much greater
distances ($>$70 pc) than the ILR, and none have sufficient N~V or C~IV
column densities to be the ILR in absorption. At least a portion of
the UV absorption component 3, at a radial velocity of $-$530 km
s$^{-1}$, is likely responsible for most of the X-ray absorption, at a
distance $<$ 7 pc from the central source. The fact that we see the ILR in
absorption in NGC~4151 and not in NGC~5548 suggests that the ILR is located
at a relatively large polar angle ($\sim$45\deg) with respect to the
narrow-line region outflow axis.

\end{abstract}

~~~~~

\section{Introduction}

About 50\% of Seyfert 1 galaxies show intrinsic absorption lines in their
UV spectra that are blueshifted with respect to their host galaxies
(Crenshaw et al. 1999; Dunn et al. 2008), indicating mass outflow from
their nuclei. The mass outflow rates are comparable to the accretion rates
needed to power the central AGN, indicating the importance of outflows in
the overall structure and energetics of active supermassive black holes
(SMBHs). Detailed studies of the absorption lines in nearby Seyfert
galaxies have shown that many are variable in their ionic column densities
over time scales of years, and, in a few cases, weeks to months (cf. Gabel
et al. 2005). The variability has been attributed to changes in the
ionizing continuum flux incident on the ``absorbers'', changes in the total
column density in the line of sight (e.g., due to transverse motion of the
absorbers against the background emission source), or a combination of the
two. In addition, changes in the fraction of the background emission that
is covered by the absorber can be interpreted as transverse motion of the
absorber (Crenshaw, Kraemer, \& Gabel 2004). In general, variability
studies provide valuable clues to the physical conditions, geometry, and
dynamics of the outflowing absorbers, and, in particular, they provide
important information on number densities, distances from the central
continuum source (presumed to be the accretion disk and corona around the
SMBH), and transverse velocities (Crenshaw, Kraemer, \& George 2003).

NGC~5548 (z $=$ 0.01676\footnote{We use this heliocentric emission-line
redshift to be consistent with our earlier papers. The NASA Extragalctic
Database gives the H~I 21-cm redshift z $=$ 0.017175 from Springob et al.
(2005). If this value is used, the radial velocities of the emission and
absorption features in the current paper should be offset by an additional
$-$125 km s$^{-1}$.}) shows multiple kinematic components of Ly$\alpha$,
N~V, and C~IV absorption in high-resolution (R $=$
$\lambda$/$\Delta\lambda$ $\geq$ 10,000) UV spectra (Crenshaw et al. 1999;
Mathur et al. 1999), at radial velocities of $-$170 to $-$1040 km s$^{-1}$
with respect to the above redshift. Multiple components of X-ray absorption
are also present in grating spectra from the {\it Chandra X-ray
Observatory} {\it CXO} (Steenbrugge et al. 2005), although the resolution
is not sufficient to identify and resolve all of the velocity components
seen in the UV. In our previous papers (Crenshaw \& Kraemer 1999, hereafter
Paper I; Crenshaw et al. 2003, hereafter Paper II), we probed the physical
conditions in the UV absorbers with spectra from the Goddard
High-Resolution Spectrograph (GHRS) and Space Telescope Imaging
Spectrograph (STIS) onboard the {\it Hubble Space Telescope} ({\it HST}).
We also presented evidence that a couple of the absorbers showed variable
ionic column densities.

In this paper, we present new {\it HST}/STIS observations
obtained in 2004 February, when the UV continuum and broad-line region
(BLR) fluxes were at an all-time low in the previous 26-year history of
satellite UV observations. It appears that NGC~5548 was in a low flux state
for several years around this time. Low X-ray fluxes were detected with
{\it CXO} in 2005 April (Detmers et al. 2008), and low BLR and continuum
fluxes were observed in the optical during 2005 March -- April (Bentz et
al. 2007). Long-term monitoring with {\it RXTE} indicates that X-ray fluxes
were in an overall low, but still variable, state during the years 2004
-- 2006 (Detmers et al. 2008).

Our recent results on the UV absorption and emission lines in another
Seyfert 1 galaxy, NGC 4151 (Crenshaw \& Kraemer 2007), illustrate the
importance of low-state observations in constraining the properties of mass
outflow in AGN. Using STIS echelle observations of NGC~4151, we identified
a component
of emission with an intermediate width (FWHM $=$ 1170 km s$^{-1}$) between
that of the BLR (FWHM $=$ 9500 km s$^{-1}$) and the narrow-line
region (NLR) (FWHM $=$ 250 km s$^{-1}$). We were able to isolate this
component by choosing a spectrum in a low-flux state, which greatly reduced
the BLR contribution to the emission, and by using a small aperture
(0\arcsecpoint2 $\times$ 0\arcsecpoint2), which significantly reduced the
NLR contribution. Based on similar velocity coverages, distances from the
continuum source, and physical conditions, we argued that the emission from
this ``intermediate-line region'' (ILR) and the broad, deep absorption in
NGC 4151, at a radial velocity of $-$490 km s$^{-1}$, arise in the same
gas. Linking the ILR emission to the broad absorption provided additional
constraints on the mass outflow in NGC~4151, such as the global covering
factor ($\sim$0.4), the outflow rate ($\sim$0.16 M$_{\odot}$ yr$^{-1}$) and
the outflow direction (centered on the accretion disk axis at polar angles
$<$ 55\deg). Additional observations of Seyfert 1 galaxies at low states
are important for testing the generality of these results, and we present
one such study here for NGC~5548.

\section{Observations}

We obtained STIS echelle spectra of the nucleus of NGC 5548 with the E140M
grating through a 0\arcsecpoint2 $\times$ 0\arcsecpoint2 aperture. We
acquired the observations over five consecutive {\it HST} orbits on each of
four consecutive days in 2004 February, for a total of 20 orbits. In Table
1, we list the details of the new observations as well as the previous {\it
HST} observations at high spectral resolution in the far-UV; the latter are
described in more detail in Papers I and II. The two GHRS observations
cover the N~V and C~IV lines separately, and were acquired $\sim$6 months
apart, when NGC~5548 was at quite different continuum levels (see Paper
II). We averaged the spectra from 2004 February, as we had done for 2002
January, because we saw no significant continuum variability ($\geq$ 3\%)
over these daily intervals. The average 2004 February spectrum has a total
exposure time of 52,156 s, which provides sufficient signal-to-noise for a
detailed analysis of this very low state. All together, we have five epochs
of UV continuum observations and four epochs of high resolution N~V and/or
C~IV observations.

In Figure 1, we show the {\it HST} high-resolution spectra in the region
around C~IV. The 2004 spectrum was obtained at a record low continuum flux
(F$_{\lambda}$[1360 \AA] $=$ 2.4 ($\pm$ 0.8) $\times$ 10$^{-15}$ ergs
s$^{-1}$ cm$^{-2}$ \AA$^{-1}$) within the 26 years of UV observations (see
the light curve in Paper II), which is $\sim$4.5 times lower than the
previous record set on 1992 July 5 (F$_{\lambda}$[1360 \AA] $=$ 1.07 ($\pm$
0.17) $\times$ 10$^{-14}$ ergs s$^{-1}$ cm$^{-2}$ \AA$^{-1}$). The broad
C~IV emission-line flux, which extends far beyond the range of the plot in
Figure 1, is also severely diminished with respect to previous
observations. Most of the remaining flux in this region is due to C~IV
emission from an intermediate-width component, with a full-width at
half-maximum (FWHM) of $\sim$700 km s$^{-1}$, compared to the widths of the
lines from the BLR and NLR. The five major kinematic components of
absorption have maintained their presence throughout the $\sim$8 years of
high-resolution UV observations\footnote{Component 5 is detected as a weak
depression in the C~IV region in 2004; it is more clearly seen as a
distinct feature in N~V (see \S3).}, despite a wide range in continuum flux
levels. We find no evidence for the appearance of new components or the
disappearance of previous components of intrinsic absorption, which we have
seen in the Seyfert 1 galaxies NGC 3783 (Kraemer, Crenshaw, \& Gabel 2001;
Gabel et al. 2005) and NGC 3516 (Kraemer et al. 2002).

\section{Emission Lines}

The intermediate-width emission in the 2004 spectrum resembles that seen in
low-state STIS echelle observations of NGC~4151 (Crenshaw \& Kraemer 2007).
We therefore used the same approach to quantify this component in NGC~5548.
We used He~II $\lambda$1640 to construct emission-line templates, because
it is a non-resonance line and is unaffected by intrinsic absorption. It
appears to have two emission components, which we show in Figure 2. We
fitted the He~II emission-line profile with two Gaussians, after
subtracting a continuum spline fit over a much larger wavelength region.
The widths of the two Gaussians are FWHM $=$ 260 and 680 km s$^{-1}$, which
we associate with the NLR (in the 0\arcsecpoint2 $\times$ 0\arcsecpoint2
aperture, which is 70 pc $\times$ 70 pc in the plane of the sky at the
distance of NGC~5548) and an intermediate-line region (ILR). By comparison,
the FWHMs
for these components in NGC~4151 are 250 and 1100 km s$^{-1}$. As can be
seen in Figure 2, the NLR template is slightly redshifted (by 120 km
s$^{-1}$) with respect to the ILR template. The BLR contribution to He~II,
which lies on the red wing of C~IV, is undetectable. Our effectiveness in
isolating the ILR component in these two Seyferts can be attributed to the
reduction of the NLR contribution by the small aperture and the reduction
of the BLR contribution due to the low flux state.

To model the emission-line profiles of the other strong lines, we used the
Gaussian profiles from the He~II fit as templates. We reproduced the
templates at the expected positions of the other strong lines (Ly$\alpha$
$\lambda$1215.670; N~V $\lambda\lambda$1238.821, 1242.804; and C~IV
$\lambda\lambda$1548.202, 1550.774), retaining the same velocity widths and
preserving the $-$120 km s$^{-1}$ offset between the NLR and ILR
components, and scaled them in intensity until we obtained suitable
matches. We allowed the doublet line ratios (e.g., C~IV
$\lambda$1548.202/$\lambda$1550.774) to vary between 1 and 2 to obtain the
best fit\footnote{Although one might expect a doublet ratio of 2, the
optically thin value, observational evidence for a ratio of $\sim$1 for the
O~VI emission line doublet in the NLR of NGC 1068 is given by Kriss et al.
(1992). We suspect that the narrow emission line doublet ratios are
affected by resonance line scattering within the outflow, and we will
explore this in a future paper.}. We detected faint broad wings in these
lines arising from the BLR, and we fitted these with cubic splines
constrained to show a single inflection at zero km s$^{-1}$ in the region
$-$2000 to $+$2000 km s$^{-1}$. We show in
Figure 3 that our method provides an excellent fit to the observed
profiles, and allows an accurate deconvolution of the NLR, ILR, and BLR
contributions to the emission lines.

From the deconvolution of the emission-line profiles, we determined the
line ratios for the narrow and intermediate-width components. Table 2 gives
these line ratios relative to He~II $\lambda$1640. Based on the
deconvolution procedure, the uncertainties in the line ratios are in the
range 10 -- 20\%. For comparison, we give the line ratios from an FOS
spectrum obtained on 1992 July 5, when NGC~5548 was in its previous very
low state (Crenshaw, Boggess, \& Wu 1993; Kraemer et al. 1998). Although
the FOS spectrum is from a larger aperture (1\arcsecpoint0 circular), and
therefore includes more NLR flux, and the FOS spectral resolution is not
sufficient to separate the narrow and intermediate-width components, the
FOS line ratios are very similar to those from the combined narrow and
intermediate components in the STIS data. The STIS intermediate-line ratios
relative to He~II are slightly higher than those from the narrow component,
but are again very similar. In Kraemer et al. (1998), we found that the NLR
was reddened by E$_{B-V}$ $\approx$ 0.07. Assuming this value holds for
both components in the STIS data, we give reddening-corrected ratios and He
II fluxes in Table 2 based on a standard Galactic reddening curve (Savage
\& Mathis 1979).

The strong C~IV line provides the best opportunity for deconvolving the
emission-line components in the other three epochs of observation, and we
did this using the same procedures and templates described above. Figure 4
shows the light curves for the C~IV emission components and the continuum
flux at 1360 \AA. It is important to note that the continuum variations are
highly undersampled over this time period; the UV continuum in NGC~5548 can
vary significantly (factor of $\sim$2) over time scales of days to weeks
(Clavel et al. 1991). Nevertheless, the overall trend from 1998 to 2004
appears to a large-scale decrease in the continuum flux in the UV and in
other wavebands, as discussed in \S1.

The NLR emission component has stayed constant to within the uncertainties
over $\sim$8 years, consistent with an origin from an extended region
(constrained by the aperture to be $\sim$70 $\times$ $\sim$70 pc in the
plane of the sky and possibly larger in the line of sight). The BLR shows
large-amplitude C~IV variations that are strongly correlated with those of
the UV continuum, consistent with previous results from more intensive UV
monitoring (Clavel et al. 1991) that the size of the C IV BLR is on the
order of light days. The ILR flux appears to shows a slight decline on a
time scales of years, consistent with a delayed response to the long-term
decline in continuum flux, indicating a size on the order of light years
for the ILR.

Due to the close agreement between the ILR line ratios and those from our
FOS observations of NGC 5548 in a low state, we can adapt our previous
photoionization models of the FOS data to model the ILR emission-line
ratios. In Kraemer et al. (1998), we required two components, INNER and
OUTER, to match the FOS line ratios. INNER is the appropriate model to use
for the ILR, since it contributes essentially all of the N~V and C~IV
emission. We have recalculated INNER using CLOUDY, version 07.02.01, last
described by Ferland et al. (1998). We used the same parameters as before
($U = 10^{-1.5}, n_H = 10^7$ cm$^{-3}$, $N_H = 10^{21.5}$ cm$^{-2}$), the
same SED as in Paper II, and roughly solar abundances (see \S5). We have
verified that the resulting line ratios agree well with our previous model
calculations. In Table 3, we show that there is good agreement between the
dereddened UV line ratios from the ILR and those from our model. The
underprediction of N~V is a common problem for photoionization models of
the NLR (see Kraemer, Bottorff, \& Crenshaw 2007 for a possible solution).

\section{Intrinsic Absorption Lines}

We list the kinematic components of absorption and their radial velocity
centroids in Table 4. In Paper II, component 6 was only detected in
Ly$\alpha$, and it is not considered in the current paper. Before we
discuss the detailed measurements of the absorption lines, we point out a
couple of interesting aspects of the features in Figure 3. First, the
maximum velocity of the absorption roughly coincides with the maximum
extent of the blue wings of the ILR profiles, which was also the case for
NGC~4151 (Crenshaw \& Kraemer 2007). Although this may just be a
coincidence, it suggests a possible connection between the absorption and
emission regions. Second, the ILR fluxes at the positions of absorption
components 2 -- 5 in Ly$\alpha$ exceed the residual fluxes in the
absorption troughs,
indicating that these absorption components at least partially cover the
ILR. Component 1, on the other hand, is not required to cover the ILR.

We have adopted two different procedures for measuring the ionic column
densities of the absorption lines. The first procedure, based on an
approach that has been used extensively in the past, including in Papers I
and II, assumes that the absorber partially covers the different background
emission components by the same fraction, except for the NLR. For the
current data, this procedure leads to serious problems with the physical
interpretation of the absorbers, as discussed in the next subsection. The
second procedure assumes different covering factors for each of the
emission components, and makes a few straightforward assumptions to derive
ionic columns that do not result in major problems in the interpretation.

\subsection{Partial Covering of All Emission Components except the NLR}

There is extensive evidence that the NLR emission in other Seyfert galaxies
is not, in general, covered by the intrinsic UV absorbers (Arav, de
Kool, \& Korista 2002; Kraemer et al.
2002; Gabel et al. 2003; Crenshaw \& Kraemer 2007). If we assume that the
NLR is uncovered and that all of the other emission components are covered
by the same fraction, then we can use the traditional doublet method
(Hamann et al. 1997) to determine the line-of-sight covering factors
($C_{f}$) and ionic columns densities of the absorption components. The
only modification to our approach in Papers I and II is that we have now
isolated the NLR and ILR emission, whereas previously we had assumed a
single emission component in the core that is partially covered. The
current approach yields essentially identical columns for our previous
data, because the NLR only contributes to the region around absorption
component 5, and in a relatively minor way (see Figure 3).

Based on the above assumptions, we can determine the covering factors from
the C~IV and/or N~V doublets from
\begin{equation}
C_f = \frac{I_1^2 - 2I_1 + 1}{I_2 - 2I_1 + 1},
\end{equation}
where I$_1$ and I$_2$ are the residual normalized fluxes in the weaker
line (e.g., C~IV $\lambda$1550.8) and stronger line (e.g.,
C~IV $\lambda$1548.2), respectively.

The optical depth at a particular wavelength is then given by
\begin{equation}
\tau = ln \left(\frac{C_f}{I_r + C_f -1}\right) 
\end{equation}

For the current data, we subtracted the NLR emission from the observed
spectrum, and normalized the absorption profiles by dividing by the sum of
the remaining emission components (continuum, BLR, ILR). We then determined
the covering factors in the troughs of the lines from equation 1. We were
unable to measure covering factors as a function of radial velocity across
the absorption lines, due to blending in the wings of the components and
because the spectrum in this very low state is still somewhat noisy despite
the long exposure time (see Figures 1 and 3). For components 1 and 5, one
member of each of the C~IV and N~V doublets is affected by blending with
another line. For these components, we assumed $C_{f}$ $=$ 1.0, as in
Paper II. For components 2, 3, and 4, our values are $C_{f}$ $=$ 0.68
($\pm$0.08), 0.63 ($\pm$0.06), and 0.84 ($\pm$0.06), respectively. These
values are slightly lower than those derived in Paper II, but agree to
within the uncertainties. We converted the normalized absorption to optical
depth profiles by using the above covering factors and integrating across
the profiles to obtain the ionic column densities (see Crenshaw et al.
2003).

In order to examine the variability of the ionic columns, we combine our
current measurements with those from Papers I and II. In Figures 5 -- 9, we
plot the continuum fluxes, ionic column densities of N~V and C~IV, and the
ratio of these columns as a function of Julian Date for each of the five
components (all indicated with a ``+'' symbol). Based on the sparse data
points, the UV continuum flux increased from 1996 to 1998, and declined
dramatically thereafter through 2002 and 2004. Component 1 showed a steady
increase in the N~V and C~IV columns from 1998 to 2004 (last 3 points), but
no evidence for a change in the N~V/C~IV ratio. Component 2 showed no
evidence of column variations outside of the error bars. Component 3 showed
significant N~V and C~IV variations, but, again, no evidence for a change
in the ratio.

Components 4 and 5 in Figures 8 and 9 show an unusual trend. The N~V/C~IV
ratios appear to have decreased significantly from 2002 to 2004,
consistent with the decrease in the UV continuum and a drop in the ionizing
flux. However, the ionic column densities of N~V and C~IV {\it dropped} as
well, except for C~IV in component 5. Based on our previous photoionization
models of these components in Paper II, we would expect that the N~V and
C~IV columns would {\it increase} with decreasing flux if the total column,
characterized by N$_H$, stayed the same. The only way to obtain the
observed N~V and C~IV columns is to have a huge decrease ($>$ factor of 10)
in N$_H$. However, these absorbers cover a significant fraction of the ILR,
which is on the order of light years in size, and a large drop in N$_H$
requires a transverse velocity on the order of the speed of light, which is
clearly unphysical, especially given the low radial velocities. Thus, we
must investigate the possibility that our initial assumptions used to
derive the column densities were incorrect.

\subsection{Different Covering Factors for Each Component}

The different emission components in our STIS aperture have very different
size scales, and there is significant evidence that UV absorbers are
typically at intermediate distances of tenths to tens of parsecs from
the central continuum source (Crenshaw et al. 2005). Thus, a reasonable
assumption is that the BLR and continuum source are fully covered,
the ILR is partially covered, and the NLR is not covered by the absorbers.
However, when we measured the ionic column densities based on this
assumption, we obtained values that were only slightly larger than those
in the previous section. This simple assumption still results in a
huge drop in N$_H$ from 2002 to 2004 for components 4 and 5, and
leaves us with unacceptably large transverse velocities.

If we examine the profiles in Figure 3, we can see that a component of
continuous emission that is constant over time, rather than a constant
fraction of the emission, might be the key to resolving the above problem.
This is a promising avenue of investigation because the relative
contribution of such a component to the total emission would be much
greater in the N~V region, where the ionic column discrepancies are much
larger than those in the C~IV region, particularly for components 4 and 5
in 2002 and 2004. The constant low-level flux may be due to starlight from
a young population (see \S6.1) or to emission from an extended scattering
region, which we have suggested for NGC~4151 (Kraemer et al. 2001).
Because we don't know {\it a priori} the exact shape of this uncovered
emission, we made the simple assumption that it is given by our simple fit
to the observed UV continuum points in 2004.

We therefore assumed that the BLR is covered, the ILR is partially covered,
and the remaining UV continuum and NLR fluxes shown in Figure 3 are
uncovered for absorption components 2 -- 5. Our main purpose is to show
that a {\it plausible} background emission model can explain the
discrepancies encountered in the previous section. For component 1, we
assumed that the BLR is covered, and the remaining emission is uncovered,
because we have no evidence for partial covering of the ILR by component 1.
Once again, we subtracted the uncovered components and normalized the
profiles by dividing by the remaining emission components. In order to
separate out the covering factors of the ILR, we further assumed that the
Ly$\alpha$ absorption is completely saturated in components 2 -- 5, which
is a very safe assumption given the strength of the Ly$\beta$ absorption in
previous {\it FUSE} observations (Brotherton et al. 2002). The residual
normalized intensities in the Ly$\alpha$ troughs of components 2 -- 5 can
be expressed as
\begin{equation}
I_r = R_b (1 - C_f^b) + R_i (1 - C_f^i),
\end{equation}
where $C_f^b$ and $C_f^i$ give the fractional covering of the BLR and ILR,
and $R_b$ and $R_i$ are the fractional contributions of the BLR and ILR to
the occulted emission (Ganguly et al. 1999, Gabel et al. 2003). Because we
are assuming that $C_f^b = 1$, the covering factor of the ILR is just
$C_f^i = (1 - I_r/R_i)$. We give the covering factors derived in this
manner in Table 4. The major effect of this new procedure on the
measurements of the ionic columns is the removal of a substantial amount of
background emission before the profiles are normalized.

The ``effective'' covering factor for each line can be determined from
\begin{equation}
C_f = R_b C_f^b + R_i C_f^i = R_b + R_i C_f^i
\end{equation}
(Gabel et al. 2003), which we used in equation 2 to determine new values
for the ionic column densities.
We give the ionic column densities derived from this approach for the 2004
spectra in Table 5. We applied the above technique to the previous GHRS and
STIS spectra, and found that the ionic columns did not change appreciably,
because the assumed unocculted continuum flux from 2004 does not contribute
significantly to the observed emission in the previous observations. We
therefore give our previously measured columns from Paper II in Table 5. We
did not detect Si~IV (or any other lower-ionization line) in the 2004
spectra and we therefore give an upper limit on its column density in the
table. Due to serious blending of the Ly$\alpha$ components, we did not
attempt to measure lower limits to the H~I column densities in the 2004
data; our previously determined limits should be sufficient.

We plot our revised column densities and ratios for the 2004 data in
Figures 5 -- 9 (indicated with an 'X' and offset by $+$50 days). For
components 4 and 5, there is no longer any evidence that the N~V/C~IV ratio
changed over the years of observation, given the uncertainties, and little,
if any, evidence that the ionic columns changed. Component 2 still shows no
evidence for changes in column density. Components 1 and 3 show a constant
N~V/C~IV ratio to within the errors, and thus no evidence for ionization
changes, but both show variable ionic column densities (as we claimed in
Paper II), indicating variations in the total hydrogen column density
(N$_H$). For component 1, a variable N$_H$ is not a problem, but rather an
indication that the absorber does not cover the ILR and that its column
density in front of the BLR has increased dramatically. Component 3 is
broad and irregular, and its structure suggests that it may consist of
several subcomponents (see Figure 1). A possible explanation for the
variability of component 3 is that it has at least one subcomponent that
does not cover the ILR and has a variable N$_H$ column, due, for example,
to bulk motion of the gas across the BLR.

\section{Photoionization Models of the Absorbers}

To explore the physical conditions in the absorbers, we calculated
photoionization models with CLOUDY to match the observed ionic column
densities from each of the STIS observations. We modeled the absorbers as
matter-bounded slabs of atomic gas, irradiated by the central course. We do
not have constraints on the relative radial positions of the absorbers, and
we therefore did not explore the effect of filtering of the ionizing
radiation (e.g. Kraemer et al. 2006). As per convention, the models are
parameterized in terms of the dimensionless ionization parameter $U$, the
ratio of the density in photons with energies $\geq$ 13.6 eV, to the number
density of hydrogen atoms at the illuminated face of the slab, and the
total hydrogen column density $N_{H}$ ($= N_{HI} + N_{HII}$), in units of
cm$^{-2}$. As mentioned in \S3, the spectral energy distribution (SED) is
detailed in Paper II. Because simultaneous X-ray observations only exist
for the 2002 dataset, we assumed the same SED for each observation.

As described in Paper II, we initially modeled the absorbers using roughly
solar abundances (e.g., Grevesse \& Anders 1989), which were by number
relative to H: He$=$0.1, C$=3.4 \times 10^{-4}$, N$=1.2 \times10^{-4}$, O
$= 6.8 \times 10^{-4}$, Ne $=1.1 \times 10^{-4}$, Mg$=3.3 \times 10^{-5}$,
Si $=3.1 \times 10^{-5}$, S$=1.5 \times 10^{-5}$, Fe$=4.0 \times 10^{-5}$.
Note that new methods for the determination of elemental abundances in the
Sun (e.g., Grevesse \& Sauval 1998), suggest that solar abundance ratios,
in particular that of nitrogen, are somewhat less than those we had
assumed. However, there is evidence that these ``old solar'' values are
approximately correct for the NLR in Seyfert galaxies (e.g. Groves, Dopita,
\& Sutherland 2004).

The N~V and C~IV columns densities and their ratios are sensitive to $U$
and $N_{H}$, and they can therefore be used to constrain the model
parameters. In Paper II, the above solar abundance models predicted larger
O~VI column densities than those determined from non-simultaneous {\it
FUSE} observations by Brotherton et al. (2002). This led us to generate
additional sets of models with higher N/C abundance ratios, which has the
effect of reducing $U$ and $N_{H}$, and, therefore, the O~VI column.
However, the {\it FUSE} observations were obtained through a very large
(30$''$ $\times$ 30$''$) aperture, when NGC 5548 was in a low state
(similar to the 1992 July observation with the FOS). If there is a
significant component of extended UV continuum emission, as we suggest in
\S4, the {\it FUSE} O~VI and Ly$\beta$ columns in Brotherton et al. (2002)
are severely underestimated, as originally found by Arav, et al. (2003),
and we can no longer use them as constraints on our models. Thus, there is
no need to include models that assume supersolar N/C.

Our approach in modeling the absorbers was as follows. Updates in the
available atomic data prompted us to rerun the models detailed in Paper II
with the latest version of Cloudy. This resulted in only slight
differences in $U$ and $N_{H}$ compared to our previous models. Our models
were deemed successful when the predicted column densities for N~V and C~IV
were within 5\% of the measured values, and the lower and upper limits for
H~I and Si~IV, respectively, were met. The one exception is the 1998 STIS
observation, for which we were only able to obtain an upper limit for C~IV;
in this case, we only required that the predicted N~V column match the
measured values to within the uncertainties. The final model parameters and
the predicted ionic column densities are given in Tables 6 and 7,
respectively.

In Table 6, the variations in $U$ and $N_H$ at different epochs reflect,
for the most part, the uncertainties in our measurements. In particular,
the components that we suggest are not variable outside of the error bars
(components 2, 4, and 5) show values of $U$ and $N_H$ that are the same to
within a factor of $\sim$2 or less. The same goes for $U$ in components 1
and 3. However, the column densities ($N_H$) in components 1 and 3 change
by as much as a factor of 10, in agreement with our claim in the previous
section.

Given the overlap in outflow velocities between the UV and X-ray absorbers
(Steenbrugge et al. 2005), it seems plausible that there is a footprint of
the X-ray absorbers in the UV. According to Table 6, the only possible
candidate for a combined UV/X-ray absorber is component 3, which has a much
higher $U$ and $N_H$ than those of the other components. In Table 8, we
compare the predicted X-ray column densities from component 3 to those
derived by Steenbrugge et al. (2005), from the contemporaneous {\it
Chandra}/HETG and LETG spectra for their Model B, for which they derived
the outflow velocities rather than fixing them to those of the UV
absorbers. Although there are several discrepancies, the overall match
is reasonable (an additional higher-ionization component would likely
eliminate some of these discrepancies). Interestingly, the
average outflow velocities derived by Steenbrugge et al. were generally in
the range of $-$400 to $-$800 km s$^{-1}$, which encompass those of UV
components 2 -- 4. Hence, it is plausible that Component 3 produced much of
the X-ray absorption detected in the {\it Chandra} spectra in 2002. 

\section{Discussion}

\subsection{Nature of the Extended UV Emission}

The two candidates for the uncovered UV continuum emission in the 2004
spectrum of NGC 5548 are: 1) scattered nuclear radiation by electrons or
dust in the inner NLR, and 2) emission from hot stars within the aperture.
The first option is not very likely, because the spectrum of the scattered
radiation should include both continuum and BLR emission. However, as shown
in Figure 3, absorption component 4 absorbs nearly all of the emission at
its position, leaving very little room for scattered broad L$\alpha$
emission. Hence, it is unlikely that the extended emission contains a
significant component of broad emission, indicating that it is not likely
due to scattered radiation from the nucleus.

Evidence for emission from hot stars in the circumnuclear region of NGC
5548 can be found in the {\it HST} F330W (U-band) image obtained on 2003
March 17 (Mu\~{n}oz Mar\'{i}n et al. 2007), which shows UV emission from a
star-forming region along an inner spiral arm $\sim$2$''$ NE of the
nucleus, as well as UV emission from other stellar clusters further away.
The inner $\sim$1$''$ of this image is dominated by the PSF of the AGN,
which we modeled with the image of a star observed in the same
configuration. We first convolved the image of NGC 5548 with a Gaussian
function to match the resolution of the PSF star. We measured the fluxes of
NGC 5548 from this convolved image and the PSF star inside concentric
circular apertures, increasing the radius in 0.5 pixel steps. Figure 10
shows the resulting radial brightness profiles of NGC 5548 and and the PSF
star normalized to the continuum peak flux. This figure also shows the
subtraction of the PSF profile, scaled by factors of 0.6, 0.75 and 0.8,
from the profile of NGC 5548. An inspection of the figure indicates that a
scaling factor between 0.75 and 0.8 produces the best residuals, by not
significantly oversubtracting the nucleus and effectively eliminating the
large PSF humps at 0.4\arcsec\ and 0.65\arcsec. Using the PSF subtracted
profiles we calculate that the 3360 \AA\ extended flux inside a region of
0\arcsecpoint2 is 1.1 $\pm0.1 \times 10^{-15}$
erg~s$^{-1}$~cm$^{-2}$~\AA$^{-1}$. This value translates to 2.5 $\pm0.2
\times 10^{-15}$ erg~s$^{-1}$~cm$^{-2}$~\AA$^{-1}$ at 1360 \AA\ for
$F_{\lambda} \propto$ $\lambda^{-1}$. This is completely consistent with
our estimate for the uncovered UV continuum emission in the STIS spectra of
F$_{\lambda}$(1360 \AA) $=$ 2.4 ($\pm$ 0.8) $\times$ 10$^{-15}$ ergs
s$^{-1}$ cm$^{-2}$ \AA$^{-1}$.

\subsection{Constraints on the Mass Outflow}

Using our photoionization models for the 2002 epoch and the lack of
evidence for a response to the UV continuum decrease from 2002 to 2004,
which provides an upper limit of 2.0 yr to the recombination time, we
constrained the electron densities in the absorbers following the method
described in Paper II, based on the equation in Bottorff, Korista, \&
Shlosman (2000). Note, we have assumed radiative recombination rates from
Shull \& van Steenberg (1982) and dielectronic recombination rates from
Nussbaumer \& Storey (1983). For components 1, 2, 4, and 5, we derive upper
limits for the electron density of $n_{e}$ $\lesssim$ 100 cm$^{-3}$ and 300
cm$^{-3}$, from C~IV and N~V, respectively. For component 3, we derived
$n_{e}$ $\lesssim$ 25 cm$^{-3}$ and 50 cm$^{-3}$, respectively. While the
model parameters for our solar models are different than the ``dusty''
models we explored in Paper II, the upper limits for $n_{e}$ are similar,
primarily due to the shorter time scale examined here.

Given $U$, an upper limit on $n_{e}$ from N V, and a luminosity of ionizing
photons corresponding to an historic mean level for NGC 5548 (Q $=$ 1.1
$\times$ 10$^{54}$ s$^{-1}$, Kraemer et al. 1998), we calculated the radial
distances $r$ of the absorbers from the ionizing continuum source. For all
the UV absorbers, r $\gtsim$ 70 pc. This limit applies to component 3 as
well, because its higher $U$ is offset by its lower $n_e$. From $U$ and
$n_H$ ($\approx$ $n_e$), we calculated a distance of $\sim$1 pc for INNER
(Kraemer et al. 1998). This is consistent with the weak response of the ILR
to continuum variations on time scales of year. The UV absorbers in NGC~
5548 are therefore at much greater distances from the central continuum
source than the inferred distance of the ILR, similar to the case for
NGC~4151, where the non-ILR absorbers span a range of distances from
1 to 2000 pc (Kraemer et al. 2001). In general, we have
found that the radial positions and velocities of most UV absorbers in
Seyfert 1 galaxies are consistent with a location in the inner NLR, at
distances of tens of pcs from the central continuum source (Crenshaw \&
Kraemer 2005).

The relatively low ionization parameter $U$ and large column $N_H$ of the
ILR, compared to the UV absorbers, result in very large columns of N~V
(10$^{16.6}$ cm$^{-2}$) and C~IV  (10$^{17.6}$ cm$^{-2}$). Thus, our line
of sight (LOS) is such that we do not see the ILR in absorption against the
central continuum source and BLR. From the ionizing continuum and H$\beta$
luminosities of NGC~5548 in
1992, we estimated that the global covering factor of INNER was $C_g
\approx$ 0.07 (Kraemer et al. 1998). Although this is
significantly smaller than the global covering factor that we derived for
NGC~4151 ($C_g \approx$ 0.4), it suggests that our chances of observing
very large column densities may be dependent on polar angle with respect to
the accretion disk axis. In fact, from kinematic studies of the NLR in
NGC~4151 based on STIS spatially-resolved spectra (Das et al. 2005), we
find that our viewing angle is $\sim$45\deg\ with respect to the outflow
axis. Although there are no similar long-slit spectra of NGC~5548, we note
that its NLR is compact and nearly circular in appearance (Schmitt et al.
2003), suggesting a smaller viewing angle.\footnote{However, we
note that the radio jet is much more extended ($\sim$18$''$) than the
NLR ($\sim$2\arcsecpoint7) in this AGN (Schmitt et al. 2003).}

Based on the above results, we present a simplified geometric picture of
the mass outflow in NGC~5548 in Figure 11. This figure is similar to one
that we presented for NGC~4151 in Kraemer et al. (2001, see Figure 6). The
major difference is that the viewing angle for NGC~4151 is significantly
larger, just outside of the NLR bicone, and therefore intercepts the ILR
(component D+E in NGC 4151, Crenshaw \& Kraemer 2007). We have identified
at least a portion of component 3 with the X-ray absorption, which Detmers
et al. (2007) place at a distance of $\leq$ 7 pc from the nucleus based on
its variability in response to a large X-ray continuum drop. The ILR, at a
distance of $\sim$1 pc, based on our photoionization models and its weak
variability on a time scale of years, is not in our line of sight. We
depict two possible explanations in Figure 11: 1) the ILR is clumpy and our
LOS does not intercept a clump, or 2) the ILR is located at large polar
angle and our LOS is at a smaller angle, consistent with the discussion in
the previous paragraph. Components 2, 4, 5, and some portion of 3 (which we
designate as 3$'$ in Figure 11) are at large distances ($>$ 70 pc), and yet
at least 4 and 5 do not completely cover the extended UV continuum source.
Component 1 is unusual in that it is at a distance $>$ 70 pc but it does
not cover the ILR, because it shows large changes in $N_H$, indicating bulk
motion across the BLR. The N~V emission from the BLR arises in a region
with a characteristic size of $\sim$4 light days (Korista et al. 1995),
which may be a lower limit to the full extent of this region. Given the
time interval between the last two observations, the transverse velocity of
component 1 is therefore v$_T$ $\geq$ 1600 km s$^{-1}$, similar to our
previously derived value in Paper 1. The same lower limit applies to the
subcomponent of 3 that changed its column density and is likely associated
with the X-ray absorber.

Based on the above parameters for the absorbers, we can calculate their
mass outflow rates from $\dot{M}_{out} = 8 \pi r N_H \mu m_p C_g v_r$,
where $m_p$ is the proton mass and $\mu$ is the mean atomic mass per
proton, assuming $C_g = 0.5$. For components 1, 2, 4, and 5, $r \geq
70$ pc and their combined mass outflow rate is $\dot{M}_{out} \geq 0.92$
M$_{\odot}$ yr$^{-1}$.\footnote{If a large portion of component 3 is at $r
< 7$ pc, then it has an upper limit of $\dot{M}_{out} < 12.0 $ M$_{\odot}$
yr$^{-1}$).} By contrast, the accretion rate into the central SMBH is
$\dot{M}_{acc} = 0.03$ M$_{\odot}$ yr$^{-1}$, based on an average
bolometric luminosity of NGC~5548 ($L_{bol} = 1.7 \times 10^{44}$ ergs
s$^{-1}$, Bentz et al. 2007) and a typical accretion disk radiative
efficiency of $\sim$0.1 in the conversion of mass infall to energy
(Peterson 1997). Thus, the outflow rate at $>$70 pc is more than 10 times
that of the SMBH accretion rate, similar to the case for NGC~4151 (Crenshaw
et al. 2007). The most probable explanation is that most of the infalling
gas does not make it to the inner accretion disk, but is instead driven
outward at these distances. The combined kinetic luminosity ($L_{KE} =
1/2\dot{M}_{out}v^2$) of the UV absorbers is $> 5.6 \times 10^{40}$ ergs
s$^{-1}$, a lower limit that is nevertheless very small compared to the
bolometric luminosity. However, we note that the above mass outflow rate
and kinetic luminosity do not include contributions from the ILR (if it is
in outflow) or the X-ray absorber.

To examine the dynamics of the outflow, we calculated the force multiplier
$FM$, which is the ratio of radiative acceleration (dominated by line
driving) to the acceleration from pure Thomson scattering, for the ionized
face and last zone of each absorber model. In order for radiative driving
to be efficient, $FM \geq (L_{bol}/L_{E})^{-1}$ (Crenshaw et al. 2003 and
references therein). NGC 5548 has a black hole mass of $M = 6.5 \times
10^{7}$ M$_{\odot}$ (Bentz et al. 2007), which yields an average Eddington
ratio of $L_{bol}/L_{E} = 0.02$. Thus the $FM$ must be $\geq$ 50 in order
for radiative driving to be important. For components 1, 2, 4, and 5, $FM
\approx 100$ at the ionized face of the cloud and $FM \approx 40$ at the
back side, so it is plausible that these components are radiatively
accelerated. For component 3, which we have linked to the X-ray absorber,
$FM = 20$ and 5 at the ionized and back sides of the absorber,
respectively. Thus, as was the case for component D+E (which was also the
principal X-ray absorber) in NGC~4151, component 3 in NGC~5548 is not
accelerated by radiation driving alone.

To investigate the possibility that component 3 is a thermal wind (Krolik
\& Kriss 1995, 2001), we calculate the radial distance at which the gas can
escape: $r_{esc} \geq G M m_H / T_g k$ (Crenshaw et al. 2003, and
references therein), where $T_g$ is the gas temperature and k is the
Boltzmann constant. Our photoionization models give $T_g =$ $1 \times 10^5$
K, which yields $r_{esc} \geq$ 340 pc. Thus, we cannot rule out a thermal
wind for the portion of component 3 at d $>$ 70 pc. However, the X-ray
absorber cannot be a thermal wind, and we suggest that magnetocentrifugal
acceleration, such as that present in an accretion-disk wind (Bottorff et
al. 2000; Everett 2005), may play an important role.

\section{Conclusions}

We have detected a distinct component of emission in a low-state {\it
HST}/STIS spectrum of NGC~5548 that has an intermediate line width with
respect to the BLR and NLR in this Seyfert 1 galaxy. The intermediate-line
region (ILR) in NGC~5548 is similar to the one that we discovered in
NGC~4151 (Crenshaw \& Kraemer 2007), although it has a somewhat larger
distance from the central continuum source ($\sim$1 pc vs. $\sim$0.1 pc),
smaller velocities (FWHM $=$ 680 km s$^{-1}$ vs. 1170 km s$^{-1}$), and a
smaller global covering factor (0.07 vs. 0.4). The source(s) of these
differences (e.g., luminosity, evolutionary effects) are not yet known, and
detection of this component in other AGN would be useful for exploring the
ILR's range in properties.

The major difference in the UV spectra of these two Seyfert 1 galaxies is
that we see the ILR in absorption in the form of broad (FWHM $\approx$ 500
km s$^{-1}$), highly-saturated absorption lines in NGC~4151, whereas we do
not see the ILR in absorption in NGC~5548. However, given the large UV
columns of the ILR in NGC~5548, its spectrum would resemble that of
NGC~4151 over a narrow range of viewing angles. Although we have no
indication that the ILR in NGC 5548 is in outflow, its resemblance to the
ILR in NGC~4151 suggests that this may be the case. Given our relatively
large viewing angle of NGC~4151 and the absence of highly-saturated UV
lines in most other Seyfert 1 galaxies (Crenshaw et al. 1999), we suggest
that ILR gas may be located at a relatively large polar angle (for Seyfert
1 galaxies) of $\sim$45\deg\ with respect to the NLR bicone (and
likely the accretion disk) axis. At even larger angles, our LOS
would presumably be blocked by the putative torus, which suggests, in
general, that the column density of absorbing material increases with polar
angle. This suggestion is supported by our imaging studies of the NLRs in
NGC~4151 (Kraemer et al. 2008) and Mrk 3 (Kraemer et al., in preparation).
More determinations of the inclinations of AGN (e.g., from the kinematics
of their NLRs) would be extremely helpful in further testing  of this
hypothesis.

\acknowledgments

Support for program 9511 was provided by NASA through a grant from the
Space Telescope Science Institute, which is operated by the Association of
Universities for Research in Astronomy, Inc., under NASA contract NAS
5-26555. Some of the data presented in this paper were obtained from the
Multimission Archive at the Space Telescope Science Institute (MAST).
SRON is supported financially by NWO, the Netherlands Organization for
Scientific Research.

\clearpage

\figcaption[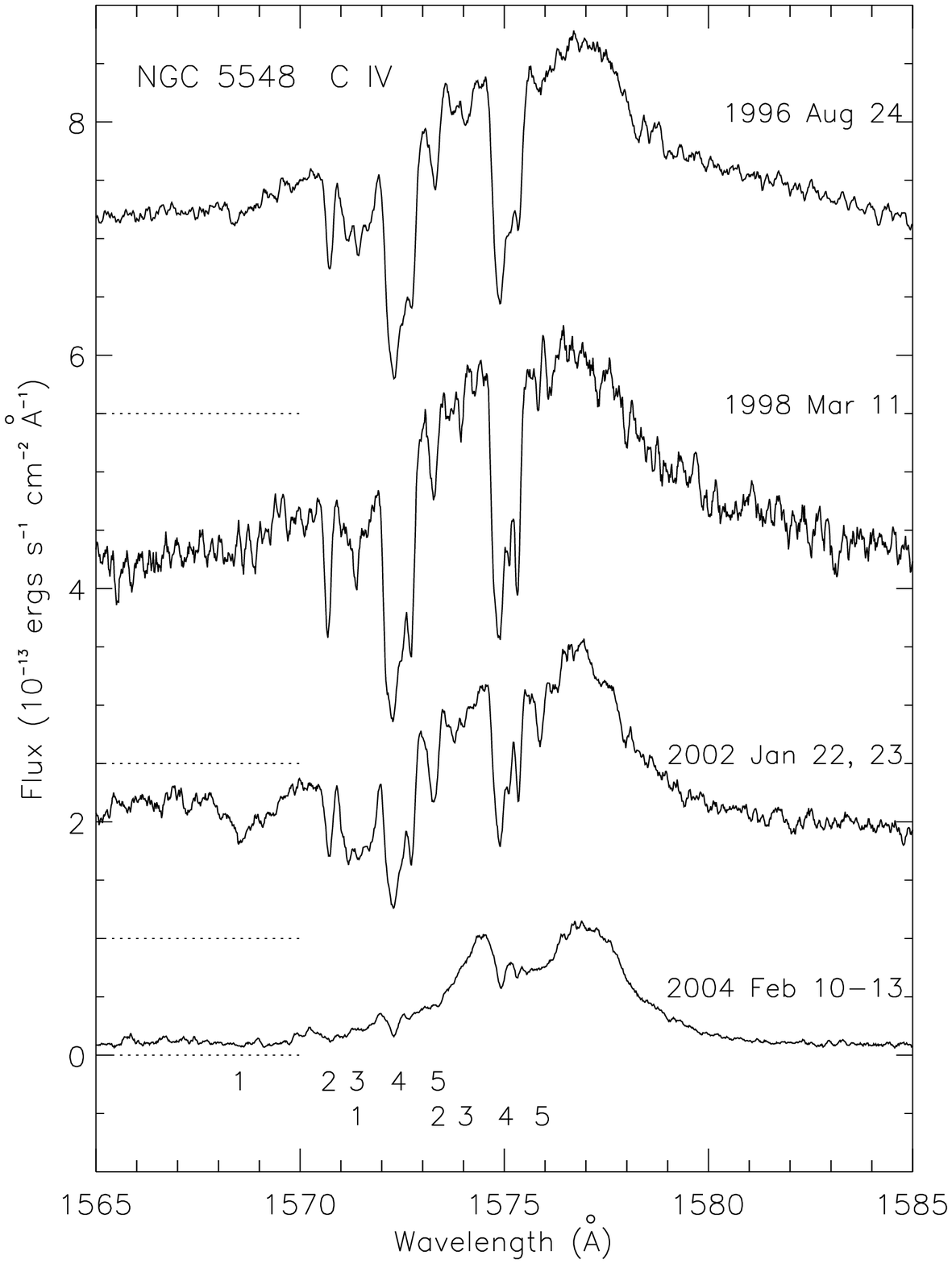]{Spectra of the C~IV region in NGC 5548 at four
epochs -- 1996 (GHRS), 1998 (STIS), 2002 (STIS) and 2004
(STIS). The UV absorption components are numbered for both members of the
doublet: C~IV $\lambda$ 1548.2 and C~IV $\lambda$ 1550.8. Dotted lines give
the zero flux level for each spectrum.}

\figcaption[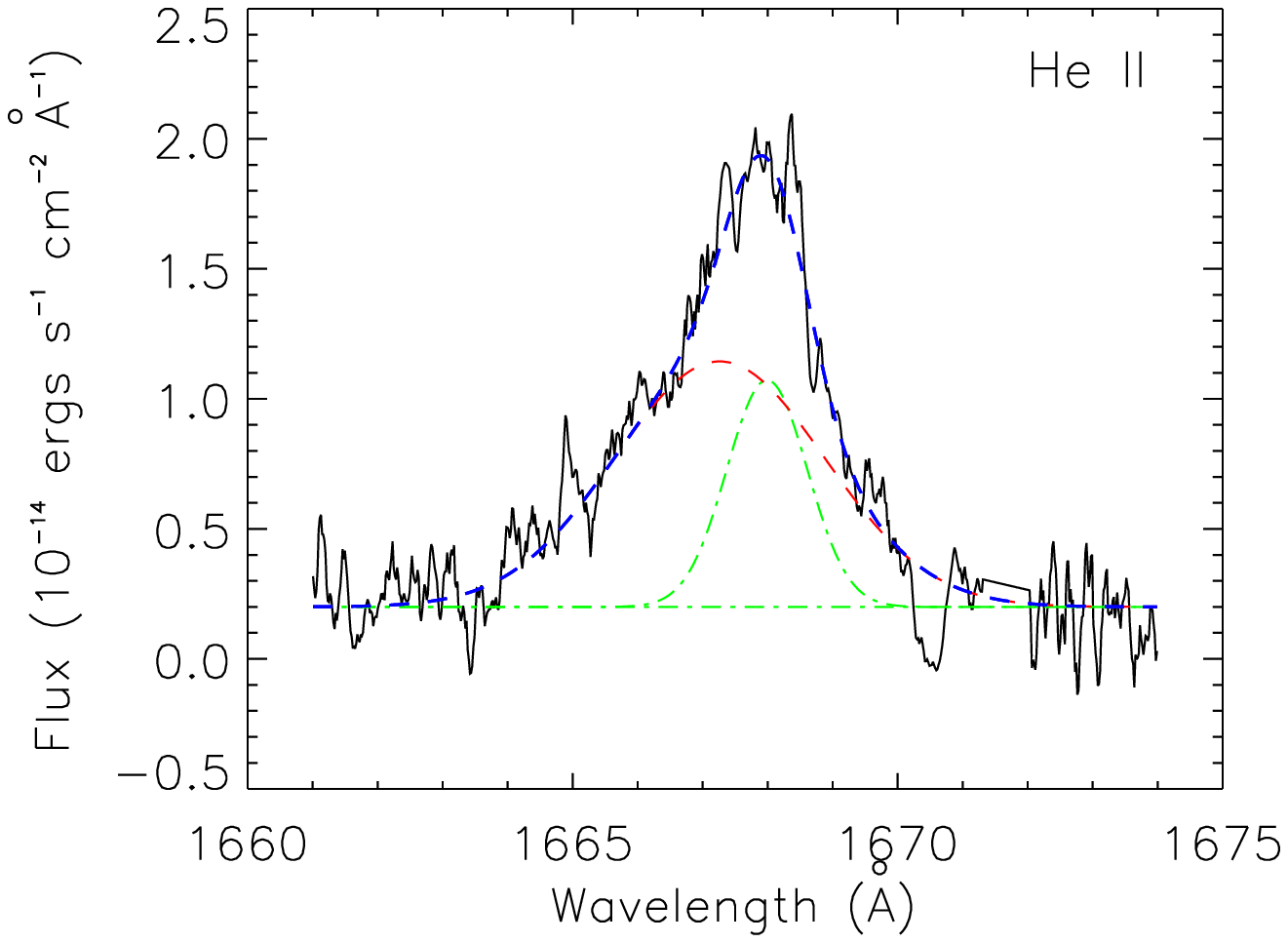]{Spectrum of the He~II $\lambda$1640 region from the
2004 STIS echelle observation of NGC 5548. Components of the emission-line
plus continuum fits are: continuum + BLR (lower dotted-dashed green),
continuum + BLR+ NLR (upper dotted-dashed green), continuum +
BLR  + ILR (dashed red) and continuum + BLR + ILR + NLR (upper dashed
blue). The absorption feature in the red wing of the He~II emission is due
to Galactic Al~II $\lambda$1670.8, which was excluded from the fits.}

\figcaption[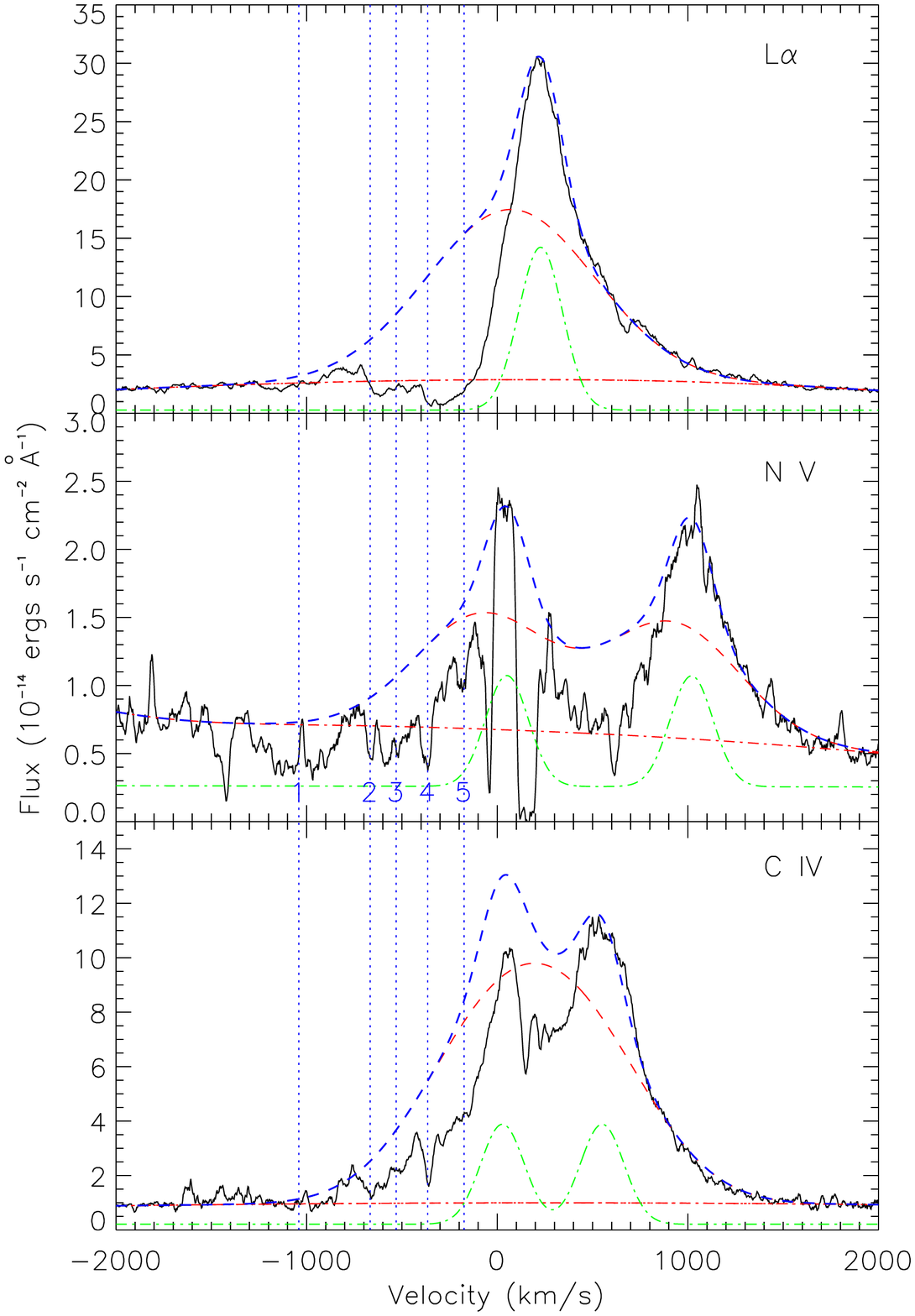]{Intrinsic absorption components and emission-line
profile fits for portions of the 2004 STIS echelle spectra of NGC~5548,
showing the intrinsic absorption lines in different ions. Fluxes are
plotted as a function of the radial velocity (of the strongest member, for
the doublets), relative to an emission-line redshift of z $=$ 0.01676. The
kinematic components are identified for the strong members of the doublets,
and vertical dotted lines are plotted at their approximate
positions. Components of the emission-line plus continuum fits
are: continuum + NLR (dotted-dashed green), continuum + BLR
(upper dotted-dashed red), continuum + BLR + ILR (dashed red) and continuum
+ BLR + ILR + NLR (upper dashed blue).}

\figcaption[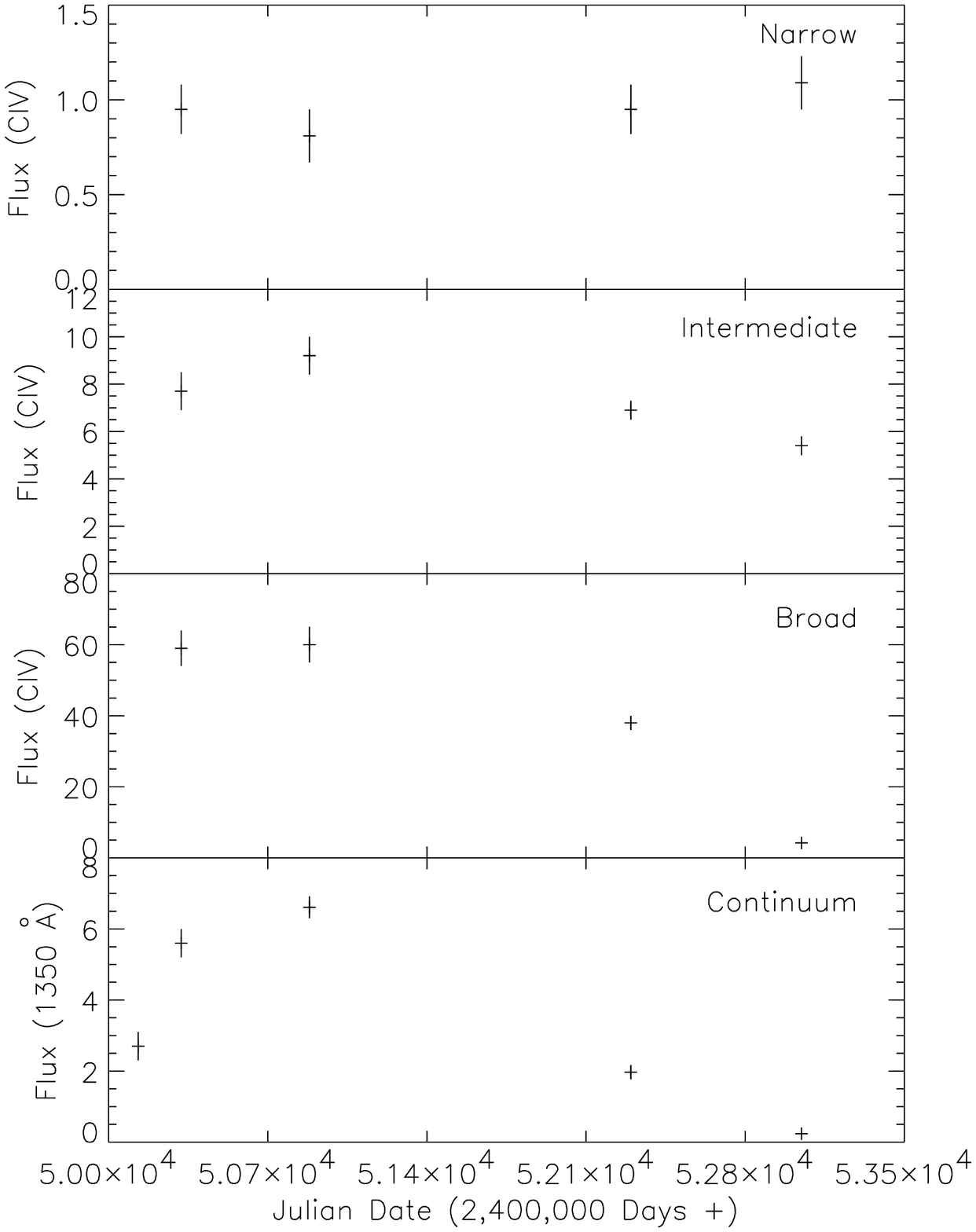]{Light curves of the continuum flux at 1360 \AA\ (bottom
panel, in units of 10$^{-14}$ ergs s$^{-1}$ cm$^{-2}$ \AA$^{-1}$) and the
C~IV emission components (top three panels, in units of 10$^{-13}$ ergs
s$^{-1}$ cm$^{-2}$).}

\figcaption[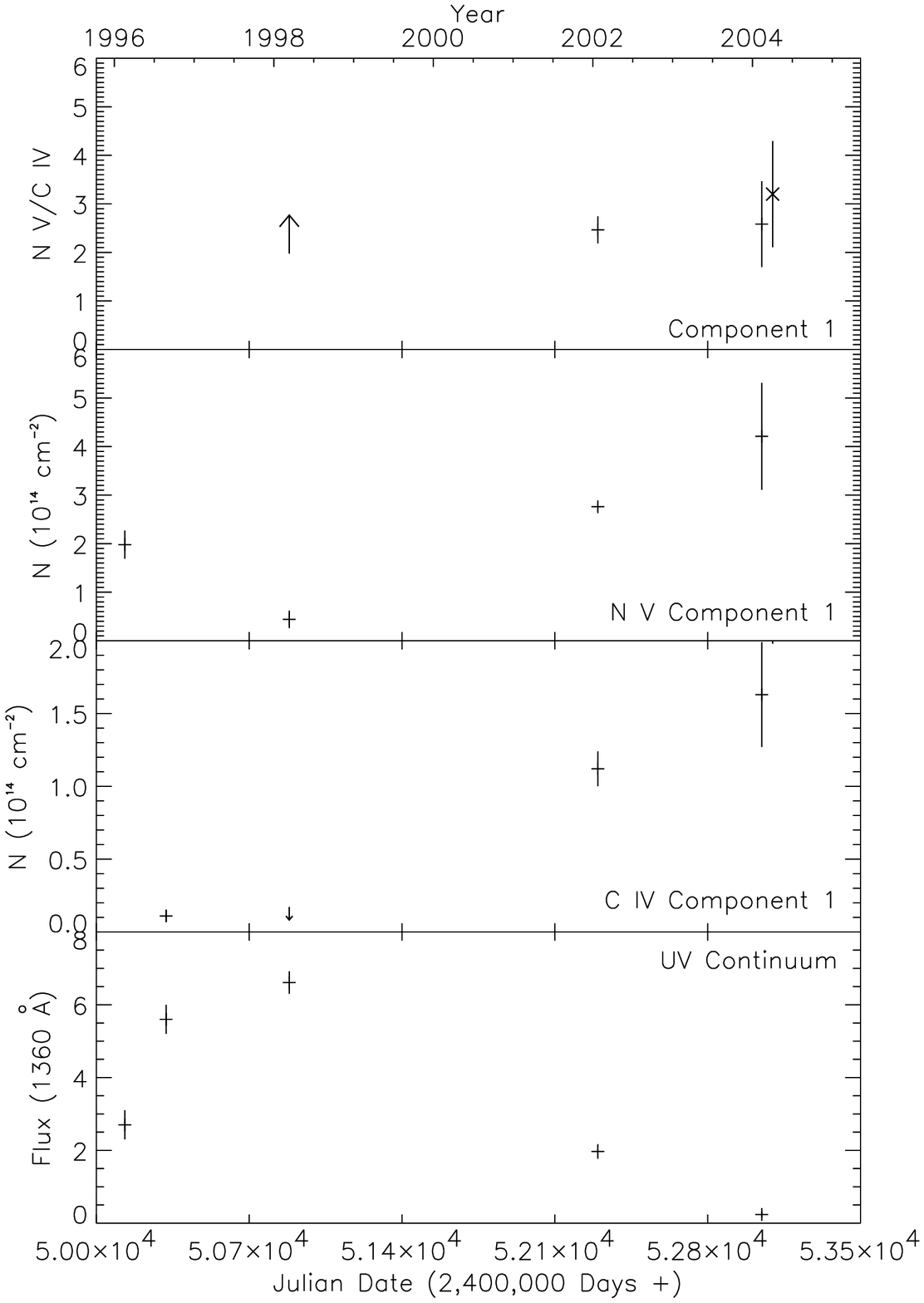]{Continuum fluxes (in units of 10$^{-14}$ ergs s$^{-1}$
cm$^{-2}$ \AA$^{-1}$), N~V and C~IV column densities, and N~V/C~IV
column density ratios
as a function of Julian Date for absorption component 1. The N~V/C~IV ratio
is not given for the first two dates, because the observations of these two
lines were not simultaneous. The original measurements of the 2004 data
are represented by a ``+'', and the revised measurements based on an
uncovered UV continuum source are given by a ``X'', offset by $+$50 days.}

\figcaption[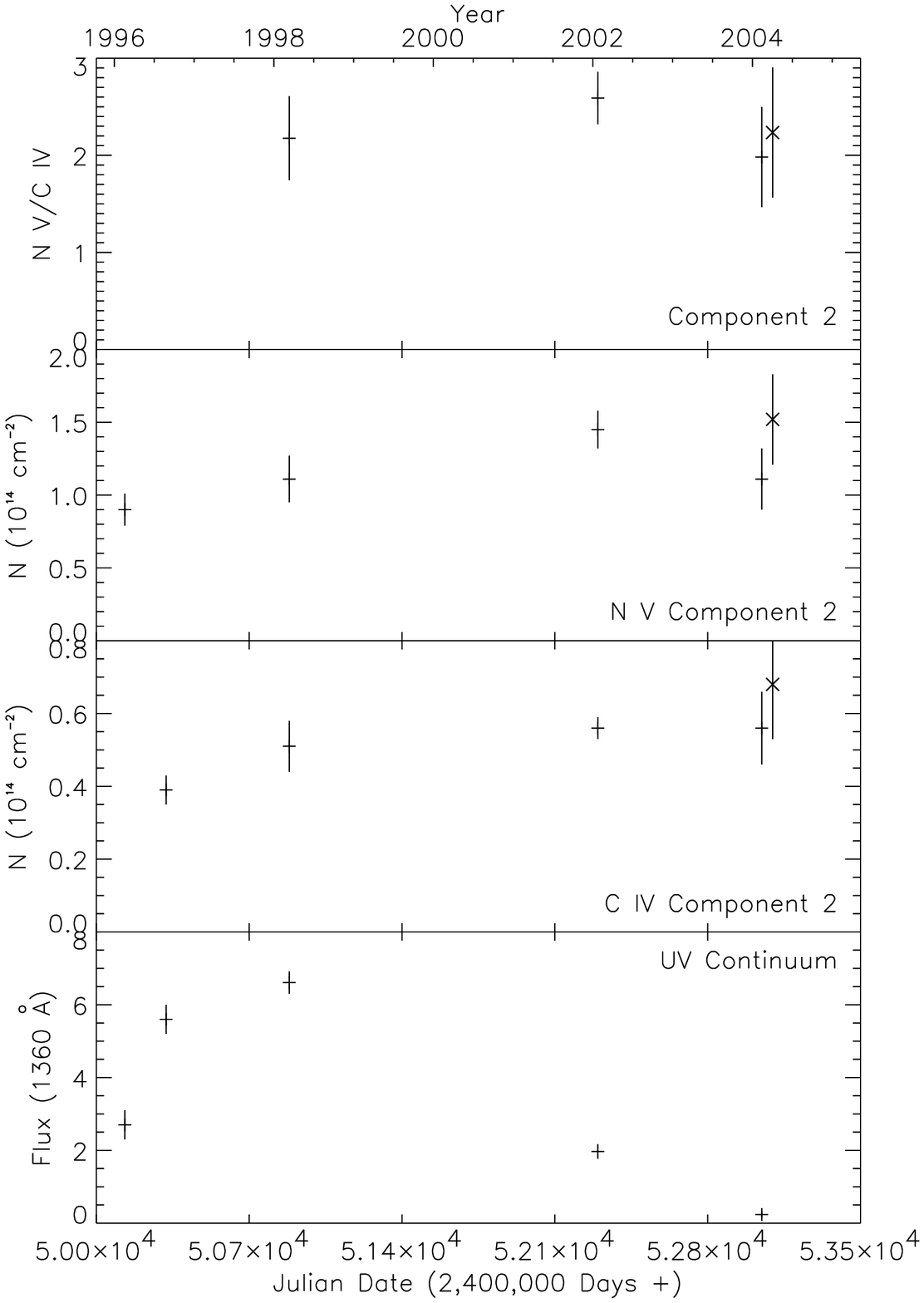]{Same as Fig. 5 for component 2.}

\figcaption[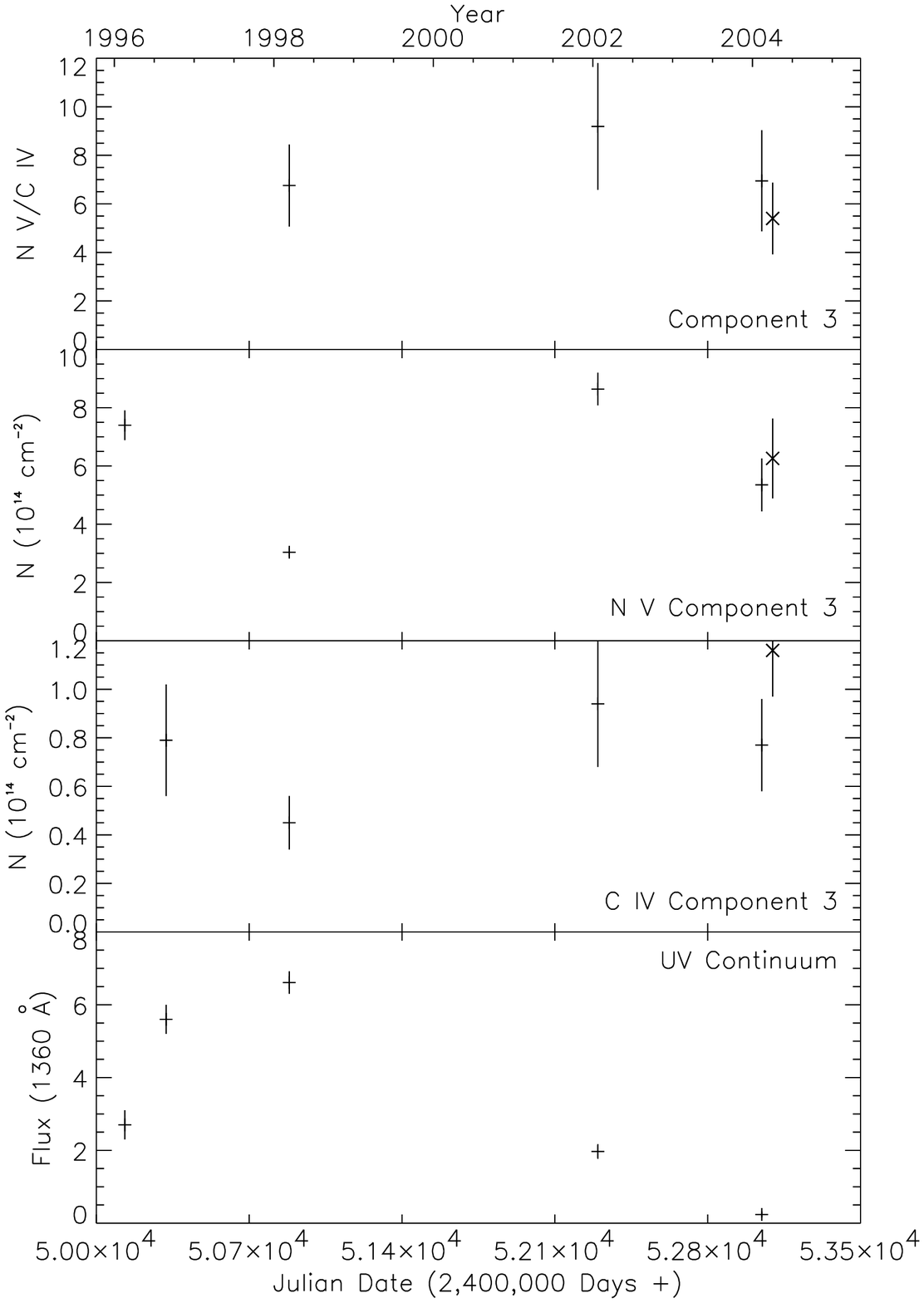]{Same as Fig. 5 for component 3.}

\figcaption[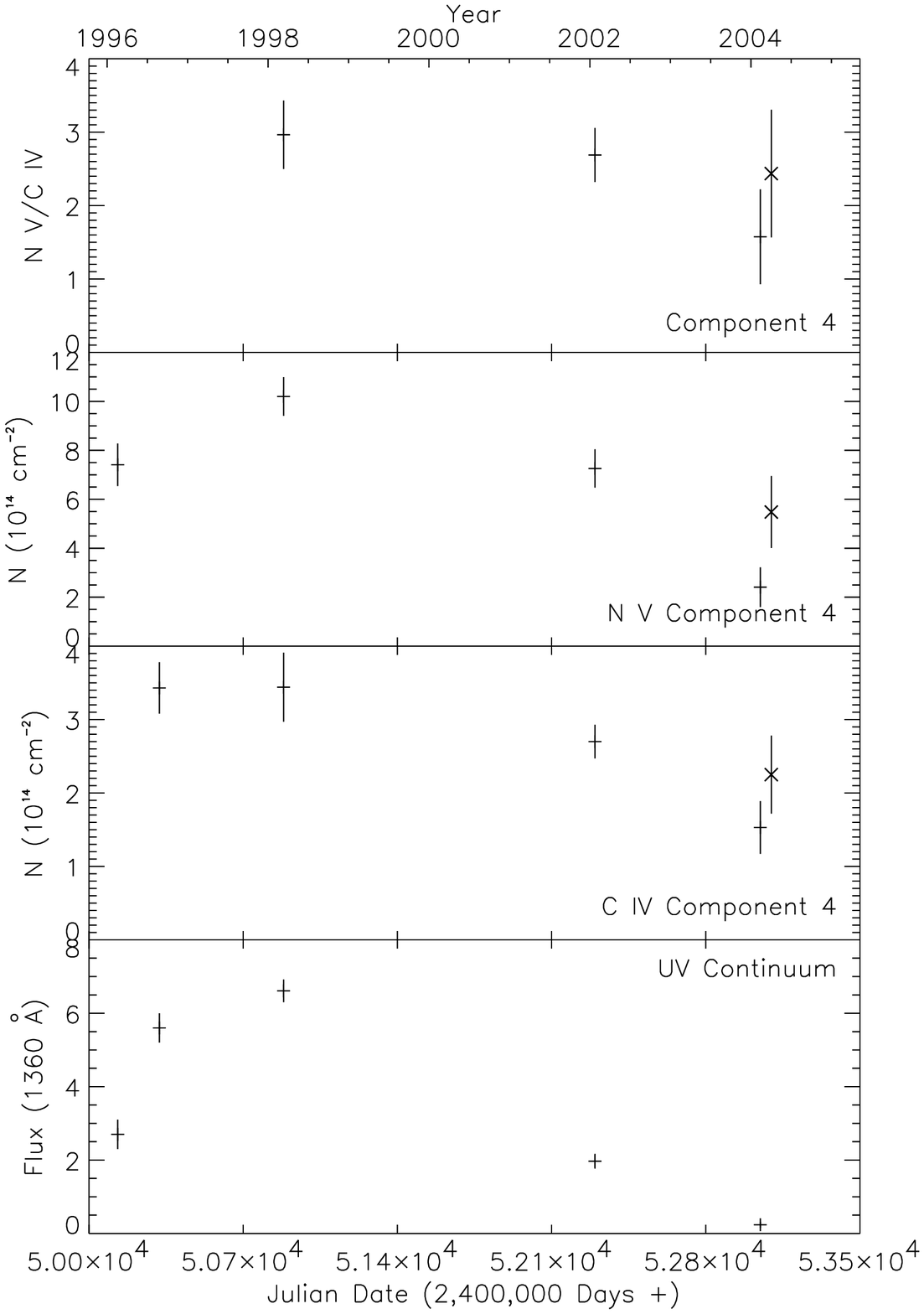]{Same as Fig. 5 for component 4.}

\figcaption[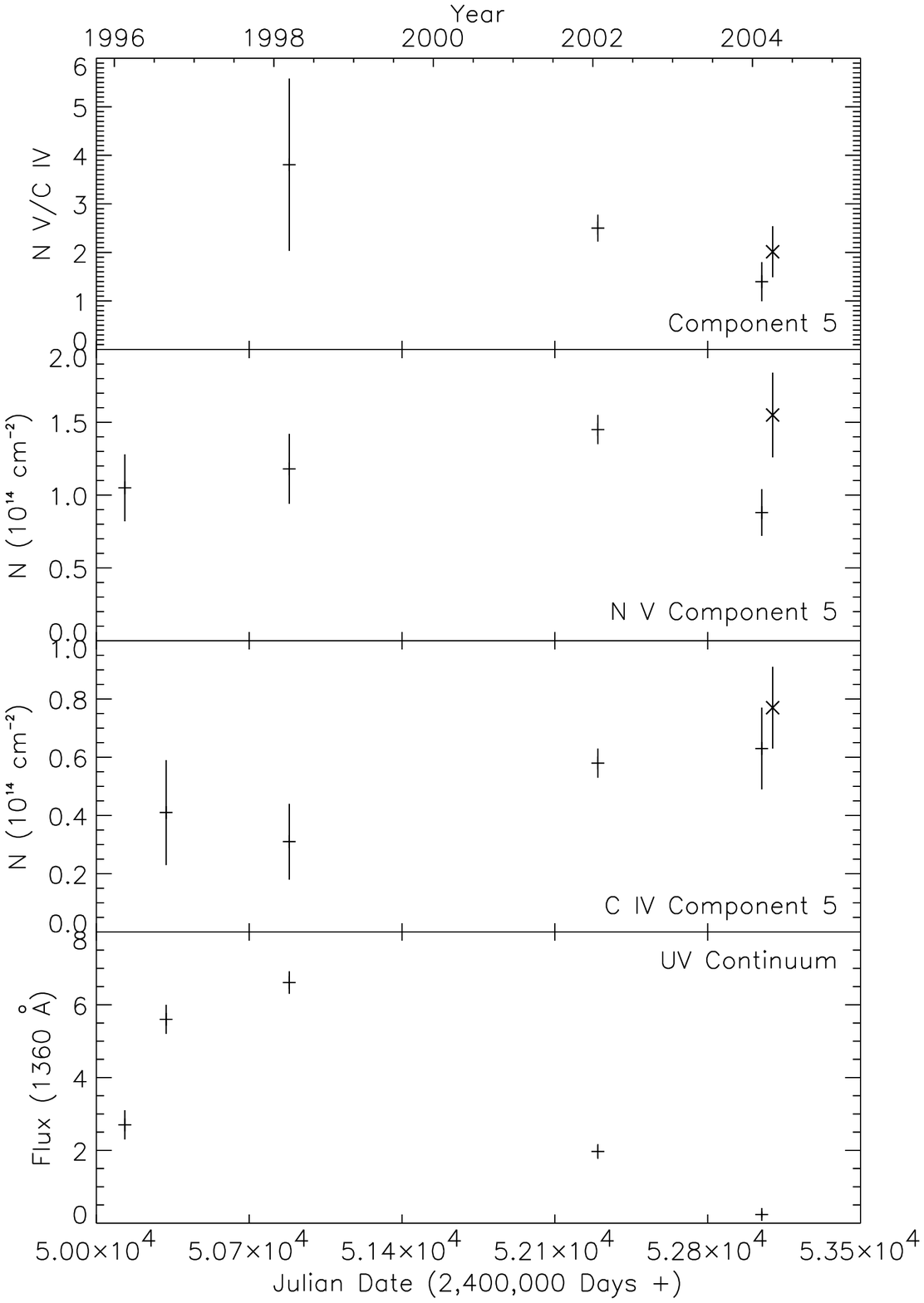]{Same as Fig. 5 for component 5.}

\figcaption[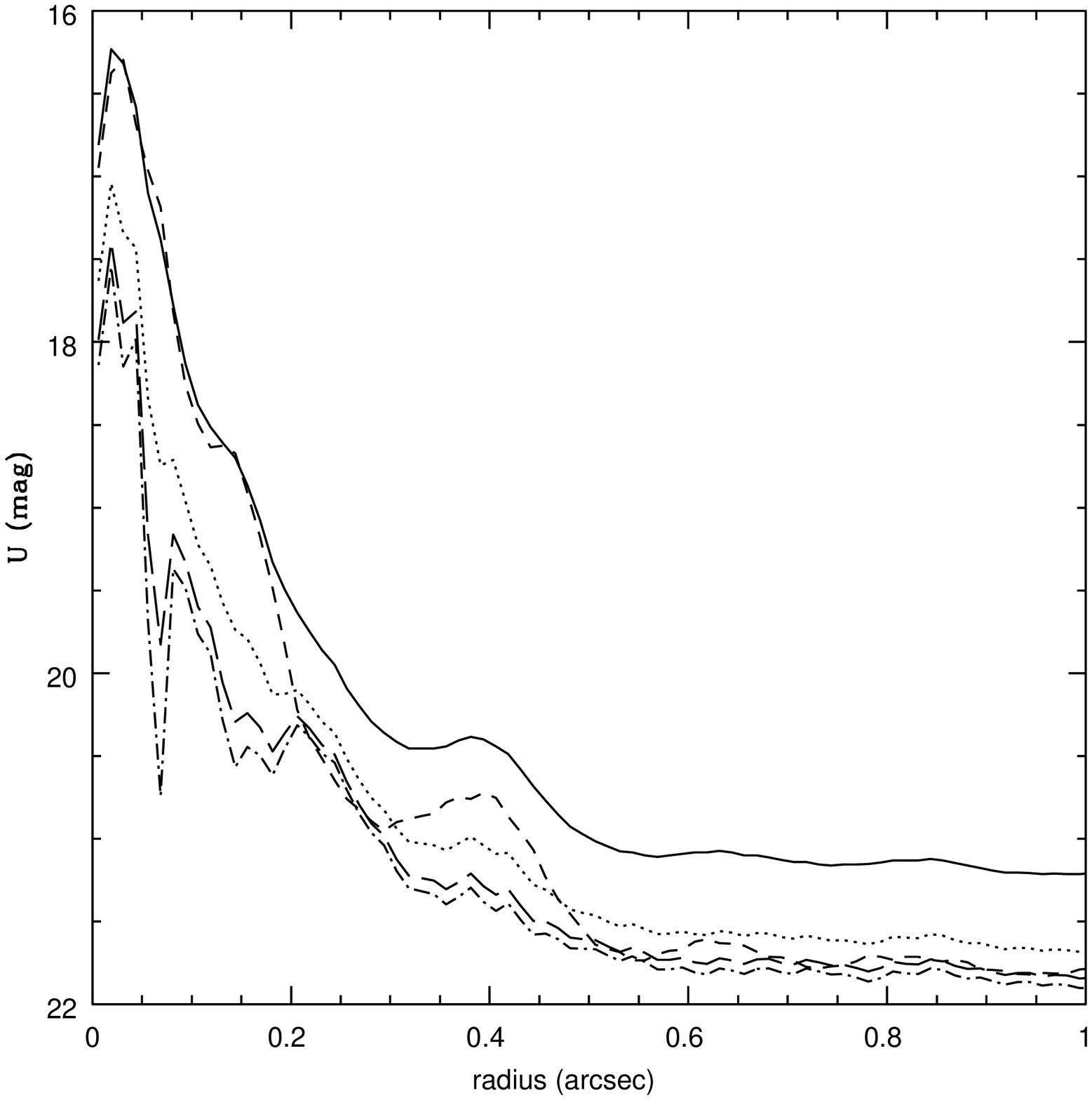]{U-band light profile of NGC~5548 (solid line),
a psf star normalized to the peak flux of NGC~5548 (short dash),
and three levels of psf subtraction from the galaxy profile. The dotted, 
long-dashed and dot-dashed lines correspond to the subtraction of a psf
normalized to 60, 75 and 80\% of the nuclear flux, respectively.}

\figcaption[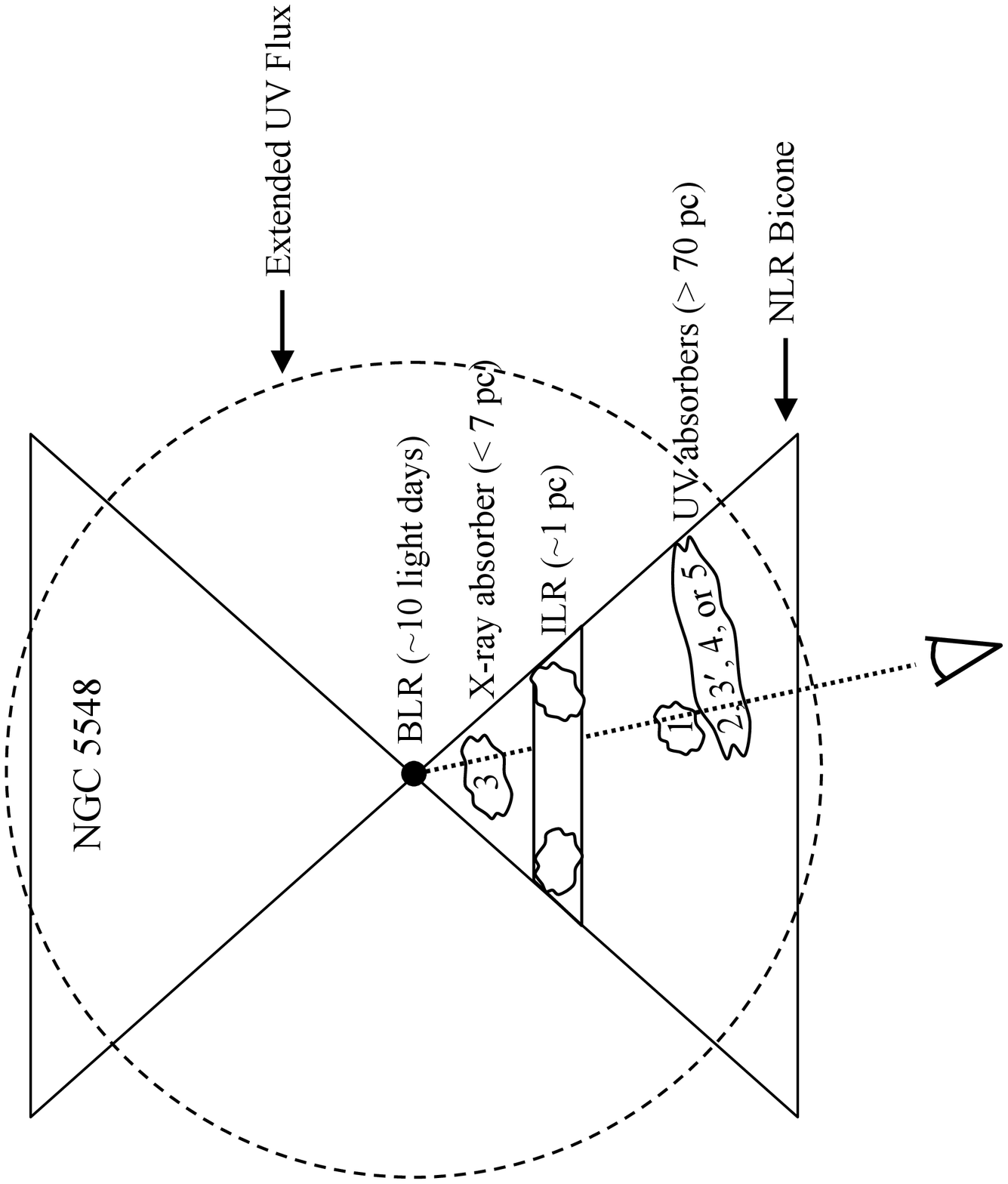]{Schematic diagram of the emission and absorption
sources in the nucleus of NGC~5548 (not to scale). Our exact line of sight
is unknown, but it is within the ionization bicone that defines the NLR
and not close to the edge of the bicone, as it is for NGC~4151.}

\newpage
\begin{deluxetable}{ccccrl}
\tablecolumns{6}
\footnotesize
\tablecaption{{\it HST} High-Resolution Spectra of NGC 5548}
\tablewidth{0pt}
\tablehead{
\colhead{Instrument} & \colhead{Grating} & \colhead{Coverage} &
\colhead{Resolution} & \colhead{Exposure} & \colhead{Date} \\
\colhead{} & \colhead{} & \colhead{(\AA)} &
\colhead{($\lambda$/$\Delta\lambda$)} & \colhead{(sec)} & \colhead{(UT)}
}
\startdata
GHRS &G160M &1232 -- 1269$^a$ &20,000 &4607   &1996 February 17 \\
GHRS &G160M &1554 -- 1590$^b$ &20,000 &13,600 &1996 August 24 \\
STIS &E140M &1150 -- 1730 &46,000 &4750   &1998 March 11 \\
STIS &E140M &1150 -- 1730 &46,000 &7639   &2002 January 23 \\
STIS &E140M &1150 -- 1730 &46,000 &7639   &2002 January 22 \\
\hline
\multicolumn{6}{c}{New Observations}\\
\hline
STIS &E140M &1150 -- 1730 &46,000 &13,039   &2004 February 10 \\
STIS &E140M &1150 -- 1730 &46,000 &13,039   &2004 February 11 \\
STIS &E140M &1150 -- 1730 &46,000 &13,039   &2004 February 12 \\
STIS &E140M &1150 -- 1730 &46,000 &13,039   &2004 February 13 \\
\enddata
\tablenotetext{a}{Covers the Ly$\alpha$ and N~V regions.}
\tablenotetext{b}{Covers the C~IV regions.}
\end{deluxetable}

\begin{deluxetable}{lrrr}
\tablecolumns{4}
\footnotesize
\tablecaption{Emission-Line Ratios (Relative to He~II $\lambda$1640)$^{a}$
\label{tbl-1}}
\tablewidth{0pt}
\tablehead{
\colhead{Component} & \colhead{Ly$\alpha$}& \colhead{N~V} & \colhead{C~IV}
}
\startdata
STIS (narrow)    &12  &1.4  &8 \\
STIS (intermediate) &17  &1.6  &14 \\
STIS (combined)  &15  &1.5  &12 \\
FOS (combined)   &12  &1.5  &9 \\
\hline
\multicolumn{4}{c}{Reddening Corrected$^{a}$}\\
\hline
STIS (narrow)$^{b}$    &14   &1.6  &8 \\
STIS (intermediate)$^{c}$ &20   &1.8  &14 \\
STIS (combined)  &17   &1.7  &12 \\
FOS (combined)   &14 &1.7  &9 \\
\hline
\enddata
\tablenotetext{a}{For E$_{B-V}$ $=$ 0.07.}
\tablenotetext{b}{He II flux $=$ 2.24 $\times$ 10$^{-14}$ ergs
s$^{-1}$ cm$^{-2}$.}
\tablenotetext{c}{He II flux $=$ 6.36 $\times$ 10$^{-14}$ ergs
s$^{-1}$ cm$^{-2}$.}

\end{deluxetable}

\begin{deluxetable}{lrrr}
\tablecolumns{4}
\footnotesize
\tablecaption{ILR Emission-Line Ratios$^a$
\label{tbl-1}}
\tablewidth{0pt}
\tablehead{
\colhead{Component} & \colhead{Ly$\alpha$}& \colhead{N~V} & \colhead{C~IV}
}
\startdata
Dereddened &20   &1.8  &14 \\
Model$^b$  &15   &0.9  &13 \\
\hline
\enddata
\tablenotetext{a}{Relative to He~II $\lambda$1640.}
\tablenotetext{b}{$U = 10^{-1.5}, n_H = 10^7$ cm$^{-3}$,
$N_H = 10^{21.5}$ cm$^{-2}$.}
\end{deluxetable}

\begin{deluxetable}{cccc}
\tablecolumns{4}
\footnotesize
\tablecaption{Absorption Components in NGC~5548}
\tablewidth{0pt}
\tablehead{
\colhead{Comp.} & \colhead{Velocity$^{a}$} & \colhead{FWHM} & 
\colhead{$C_f^i$ $^{b}$}\\
\colhead{} &\colhead{(km s$^{-1}$)} &\colhead{(km s$^{-1}$)} & \colhead{}
}
\startdata
1        &$-$1040   & 220  & --- \\
2        &$-$670    & 40   & 0.71 ($\pm$0.04)  \\
3        &$-$530    & 160  & 0.76 ($\pm$0.03) \\
4        &$-$340    & 150  & 0.96 ($\pm$0.01) \\
5        &$-$170    & 60   & 0.75 ($\pm$0.04)\\
\enddata
\tablenotetext{a}{Velocity centroid for a systemic redshift of z $=$
0.01676.}
\tablenotetext{b}{Covering fraction of the ILR, based on the uncovered
continuum case.}
\end{deluxetable}

\begin{deluxetable}{clcccc}
\tablecolumns{6}
\footnotesize
\tablecaption{Measured Ionic Column Densities (10$^{14}$ cm$^{-2}$)}
\tablewidth{0pt}
\tablehead{
\colhead{Comp.} & \colhead{Ion} & \colhead{GHRS$^a$} & 
\colhead{STIS 1998} & \colhead{STIS 2002} & \colhead{STIS 2004$^b$}
}
\startdata
1     &H~I     &---        &$>$1.34       &$>$2.31 &---\\
      &N V     &1.98 (0.29)   &0.44 (0.18)   &2.76 (0.10) &8.13 (2.12)\\
      &C IV    &0.11 (0.04)  &$<$0.17         &1.12 (0.12) &2.54 (0.56)\\
      &Si IV   &---       &$<$0.10        &$<$0.06 &$<$0.22\\
& & & & \\
2     &H~I      &$>$0.76        &$>$0.61   &$>$1.18 &---\\
      &N V     &0.90 (0.11)   &1.11 (0.16)   &1.45 (0.13) &1.52 (0.31)\\
      &C IV    &0.39 (0.04)   &0.51 (0.07)   &0.56 (0.03) &0.68 (0.15)\\
      &Si IV   &---       &$<$0.10        &$<$0.06 &$<$0.22\\
& & & & \\
3     &H~I      &$>$1.79        &$>$2.01   &$>$1.65 &---\\
      &N V     &7.40 (0.51)   &3.04 (0.16)   &8.64 (0.56) &6.26 (1.37)\\
      &C IV    &0.79 (0.23)   &0.45 (0.11)   &0.94 (0.26) &1.16 (0.19)\\
      &Si IV   &---        &$<$0.10        &$<$0.06 &$<$0.22\\
& & & & \\
4     &H~I     &$>$3.30        &$>$2.97 &$>$3.64 &---\\
      &N V     &7.41 (0.87)  &10.23 (0.79)   &7.26 (0.78) &5.48 (1.47)\\
      &C IV    &3.43 (0.35)   &3.44 (0.47)   &2.70 (0.23) &2.25 (0.53)\\
      &Si IV   &---        &$<$0.10        &$<$0.06 &$<$0.22\\
& & & & \\
5     &H~I      &$>$0.77     &$>$0.66  &$>$0.92 &---\\
      &N V     &1.05 (0.23)   &1.18 (0.24)    &1.45 (0.10) &1.55 (0.29)\\
      &C IV    &0.41 (0.18)   &0.31 (0.13)   & 0.58 (0.05) &0.77 (0.14)\\
      &Si IV   &---         &$<$0.10         &$<$0.06 &$<$0.22\\
\enddata
\tablenotetext{a}{GHRS C IV is not simultaneous with GHRS N V and H I.}
\tablenotetext{b}{Based on the uncovered continuum case (see text)}
\end{deluxetable}

\begin{deluxetable}{lccc}
\tablecolumns{5}
\footnotesize
\tablecaption{Model Parameters}
\tablewidth{0pt}
\tablehead{
\colhead{Observation} &\colhead{Component} &\colhead{log U} & \colhead{log
N$_{H}$} 
}
\startdata
STIS 1998 & & &  \\ 
& 1  & $-0.50$ & 19.23\\
& 2 &  $-0.84$ & 19.03\\
& 3 & $-0.12$ & 21.15\\
& 4 & $-0.60$ & 20.35\\ 
& 5 &  $-0.48$ & 19.66\\
STIS2002 & & &  \\
& 1 & $-0.74$ & 19.55\\
& 2 &  $-0.71$ & 19.34\\
& 3 &  0.10 & 22.2 \\
& 4  & $-0.68$ & 20.06 \\
& 5 &  $-0.72$ & 19.30\\
STIS2004  & & & \\
 & 1 &  $-0.55$ & 20.35\\
& 2 &  $-0.80$ & 19.2 \\
& 3 &  $-0.24$ & 21.02\\
& 4 &  $-0.74$ & 19.85\\
& 5 & $-0.90$ & 19.07 \\                    
\enddata
\end{deluxetable}

\begin{deluxetable}{clccc}
\tablecolumns{5}
\footnotesize
\tablecaption{Predicted Column Densities(10$^{14}$ cm$^{-2}$)$^{a}$}
\tablewidth{0pt}
\tablehead{
\colhead{Comp.} & \colhead{Ion} & 
\colhead{STIS 1998} & \colhead{STIS 2002} & \colhead{STIS 2004}
}
\startdata
1     &H~I      &1.52       &6.41 & 22.0\\
      &N V      &0.50   &2.73 &8.11\\
      &C IV     &0.14         &1.11 &2.48\\
 & & & \\
2     &H~I              &2.58   &3.60 & 3.43\\
      &N V        &1.14    &1.50  & 1.50\\
      &C IV       &0.53    &0.58  & 0.66\\
&  & & \\      
3     &H~I       &23.9   &88.9 &31.0\\
       &N V     &3.01   &8.51 &6.21\\
       &C IV     &0.44   &0.98 &1.12 \\
& & & \\
4     &H~I     &25.9 &17.2 & 12.8\\
      &N V       &10.21    &7.19  & 5.53\\
      &C IV    &3.39      &2.69  & 2.24\\
& &  &  \\
5     &H~I      &3.50  & 3.38 & 3.43\\
      &N V       &1.13     &1.42  & 1.53\\
      &C IV      &0.31    & 0.56  & 0.76\\
\enddata
\tablenotetext{a}{Si~IV column densities for all models were $<$ 10$^{12}$
cm$^{-2}$.}
\end{deluxetable}

\begin{deluxetable}{lcc}
\tablecolumns{3}
\footnotesize
\tablecaption{Predicted$^{a}$ and Measured$^{b}$ X-ray Column Densities for
2002 spectra}
\tablewidth{0pt}
\tablehead{
\colhead{ion} &\colhead{Predicted logN$_{i}$} &\colhead{Measured
logN$_{i}$}  
}
\startdata
C~V & 16.9 & 17.2\\
C~VI & 18.1 & 17.6\\
N~VI & 17.2 & 17.2\\
N~VII & 18.0 & 17.5 \\
O~V & 15.7 & 16.8 \\
O~VI & 16.8 & 16.6 \\
O~VII & 18.5 & 18.2 \\
O~VIII & 18.7 & 18.5 \\
Ne~IX & 18.0 & 17.2 \\
Ne~X & 17.6 & 17.9 \\
Mg~XI & 17.3 & 17.0\\
Mg~XII & 16.5 & 17.3\\
Si~VIII & 16.7 & 16.2\\
Si~X & 17.2 & 16.0\\
Si~XI & 16.8 & 17.0 \\
Si~XIII & 16.4 & 17.0\\
Si~XIV & 15.2 & 17.0\\
S~XI & 16.9 & 16.0\\
S~XII & 16.7 & 16.2 \\
Fe~XVII & 15.1 & 16.3 \\
\enddata
\tablenotetext{a}{Predicted column densities for component 3 in the   
STIS 2002 spectra, assuming solar abundances.}
\tablenotetext{b}{Measured columns for Model B of Steenbrugge et al.
(2005).}
\end{deluxetable}

\clearpage
\begin{figure}
\plotone{f1.eps}
\\Fig.~1.
\end{figure}

\clearpage
\begin{figure}
\plotone{f2.eps}
\\Fig.~2.
\end{figure}

\clearpage
\begin{figure}
\plotone{f3.eps}
\\Fig.~3.
\end{figure}

\clearpage
\begin{figure}
\plotone{f4.eps}
\\Fig.~4.
\end{figure}

\clearpage
\begin{figure}
\plotone{f5.eps}
\\Fig.~5.
\end{figure}

\clearpage
\begin{figure}
\plotone{f6.eps}
\\Fig.~6.
\end{figure}

\clearpage
\begin{figure}
\plotone{f7.eps}
\\Fig.~7.
\end{figure}

\clearpage
\begin{figure}
\plotone{f8.eps}
\\Fig.~8.
\end{figure}

\clearpage
\begin{figure}
\plotone{f9.eps}
\\Fig.~9.
\end{figure}

\clearpage
\begin{figure}
\plotone{f10.eps}
\\Fig.~10.
\end{figure}

\clearpage
\begin{figure}
\plotone{f11.eps}
\\Fig.~11.
\end{figure}


\clearpage			       
\begin{references}

\reference{ara2002}Arav, N., Korista, K.T., \& de Kool, M. 2002, \apj,
566, 699

\reference{arav2003}Arav, N., et al. 2003, \apj, 590, 174

\reference{ben2007}Bentz, M.C., et al. 2007, \apj, 662, 205

\reference{bot2000}Bottorff, M.C., Korista, K.T., \& Sholsman, I. 2000,
\apj, 537, 2000

\reference{bro2002}Brotherton, M.C., Green, R.F., Kriss, G.A., Oegerle, W.,
Kaiser, M.E., Zheng, W., \& Hutchings, J.B. 2002, \apj, 565, 800 

\reference{cla1991}Clavel, J., et al. 1991, \apj, 366, 64

\reference{cre1993}Crenshaw, D.M., Boggess, A., \& Wu, C.-C. 1993, \apj,
416, L67

\reference{cre1999}Crenshaw, D.M., \& Kraemer, S.B. 1999, \apj, 521, 572
(Paper I)

\reference{cre2004}Crenshaw, D.M., \& Kraemer, S.B. 2005, \apj, 625, 680

\reference{cre2007}Crenshaw, D.M., \& Kraemer, S.B. 2007, \apj, 659, 250

\reference{cre1999}Crenshaw, D.M., Kraemer, S.B., Boggess, A., Maran, S.P.,
Mushotzky, R.F., \& Wu, C.-C. 1999, \apj, 516, 750.

\reference{cre2001}Crenshaw, D.M., Kraemer, S.B., \& Gabel, J.R. 2001,
\apj, 557, 30

\reference{cre2004}Crenshaw, D.M., Kraemer, S.B., \& Gabel, J.R. 2004,
in AGN Physics with the Sloan Digital Sky Survey, ed. G.T. Richards \&
P.B. Hall (San Francisco: Astronomical Society of the Pacific), ASP
Conference Series, 311, 235

\reference{cre2003a}Crenshaw, D.M., Kraemer, S.B., \& George, I.M. 2003,
\araa, 41, 117

\reference{cre2003b}Crenshaw, D.M., et al. 2003, \apj, 594, 116 (Paper II)

\reference{das2005}Das, V., et al. 2005, \aj, 130, 945.

\reference{det2008}Detmers, R.G., Kaastra, J.S., Constantini, E., McHardy,
I.M., \& Verbunt, F. 2008, A\&A, 488, 67

\reference{dun2008}Dunn, J.P., Crenshaw, D.M., Kraemer, S.B., \& Trippe,
M.L. 2008, \aj, 136, 1201

\reference{eve2005}Everett, J.E. 2005, \apj, 631, 689.

\reference{fer1998}Ferland, G.J, et al. 1998, PASP, 110, 761


\reference{gan1999}Ganguly, R., Eracleous, M., Charlton, J,C., \&
Churchill, C.W. 1999, AJ, 117, 2594


\reference{gab2003}Gabel, J.R., et al. 2003, \apj, 583, 178

\reference{gab2005}Gabel, J.R., et al. 2005, \apj, 631, 741

\reference{gre1989}Grevesse, N., \& Anders, E. 1989, in AIP Conf. Proc.
183, Cosmic 
Abundances of Matter, ed. C.J. Waddington (New York:AIP), 1

\reference{gre1998}Grevesse, N., \& Sauval, A.J. 1998, Space Sci. Rev., 85,
161

\reference{gro2004}Groves, B.A., Dopita, M.A., \& Sutherland, R.S. 2004,
ApJS,
153, 9

\reference{ham1997}Hamann, F., Barlow, T.A., Junkkarinen, V., \& Burbidge,
E.M. 1997, \apj, 478, 78

\reference{kor1995}Korista, K.T., et al. 1995, \apjs, 97, 285

\reference{kra2007}Kraemer, S.B., Bottorff, M.,C., \& Crenshaw, D.M. 2007,
\apj, 668, 730

\reference{kra1998}Kraemer, S.B., Crenshaw, D.M., Filippenko, A.V., \&
Peterson, B.M. 1998, \apj, 499, 719

\reference{kra2002}Kraemer, S.B., Crenshaw, D.M., George, I.M., Netzer,
H., Turner, T.J., \& Gabel, J.R. 2002, \apj, 577, 98

\reference{kra2008}Kraemer, S.B., Schmitt, H.R., \& Crenshaw, D.M. 2008
\apj, 679, 1128

\reference{kra2001}Kraemer, S.B., et al. 2001, \apj, 551, 671.

\reference{kra2006}Kraemer, S.B. et al. 2006, \apjs, 167, 161

\reference{kri1992}Kriss, G.A., et al. 1992, \apj, 392, 485

\reference{kro1995}Krolik, J.H. \& Kriss, G.A. 1995, \apj, 447, 512

\reference{kro2001}Krolik, J.H. \& Kriss, G.A. 2001, \apj, 561, 684

\reference{mat1999}Mathur, S., Elvis, M., \& Wilkes, B.J. 1999, \apj, 519,
605

\reference{mun2007}Mu\~{n}oz Mar\'{i}n, V.M., et al. 2007, \aj, 134, 648

\reference{nus1983}Nussbaumer, H., \& Storey, P.J. 1983, A\&A, 126, 75

\reference{pet1997}Peterson, B.M. 1997, An Introduction to Active Galactic
Nuclei (Cambridge, UK: Cambridge University Press).

\reference{sav1979}Savage, B.D., \& Mathis, J.S. 1979, ARAA, 17, 73

\reference{sch2003}Schmitt, H.R., Donley, J.L., Antonucci, R.R.J.,
Hutchings, J.B., \& Kinney, A.L. 2003, \apjs, 148, 327

\reference{shu1982}Shull, J.M., \& van Steenberg, M. 1982, ApJS, 48, 95

\reference{ste2005}Steenbrugge, K. C., et al. 2005, A\&A, 434, 569

\end{references}
\end{document}